\documentclass[preprint]{revtex4}
\usepackage{amsfonts,amssymb,epsfig}
\begin{document}
\sloppy
\newcommand{\fdag}{\not\kern-.25em}

\title{
The NN Phase Shifts $<3$ GeV and Resonance Features
\footnote{To appear in the Procedings of the International Workshop on
Resonances in Few-Body Systems,  4-8 September 2000, Sarospatak Hungary,
Few-Body Systems, Supplement, Springer Verlag }
}
\author{Funk, A.}
\author{von Geramb, H.V. }
\affiliation{
Theoretische Kernphysik, Universit\"at Hamburg, 
Luruper Chaussee 149, D-22761 Hamburg, Germany
}

\begin{abstract}
Solutions SP00, SM00 and FA00 of nucleon-nucleon (NN) phase shift analyzes by
Arndt {\em et al.} are used in optical model studies. The partial waves, 
single channels $J\leq7$ and  coupled channels $J\leq6$, are scrutinized. 
The radial probability distributions and losses of flux are used to identify 
the known $\Delta$ and $N^*$ resonances as well as anticipated other structures.
The energy  interval extends to 3 GeV.
\end{abstract}

\maketitle

\section{Introduction and Salient Features}

\begin{figure}[ht]
\centering
\epsfig{file=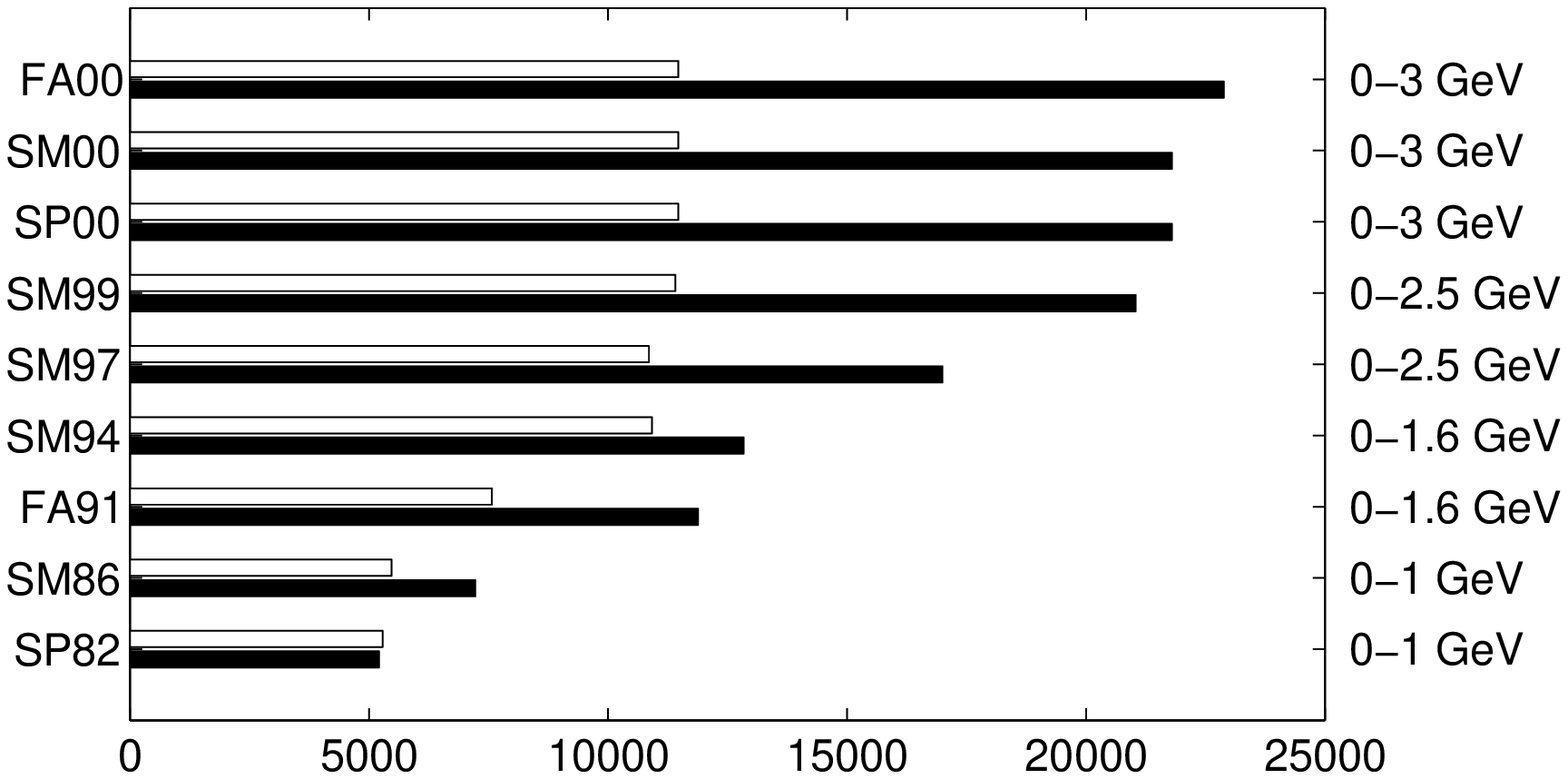,scale=.4}
\caption{History of database SAID  1980-2000,
solid bars are pp data and open bars are np data}
\label{arnsol1}
\end{figure}

\begin{figure}[ht]
\centering
\epsfig{file=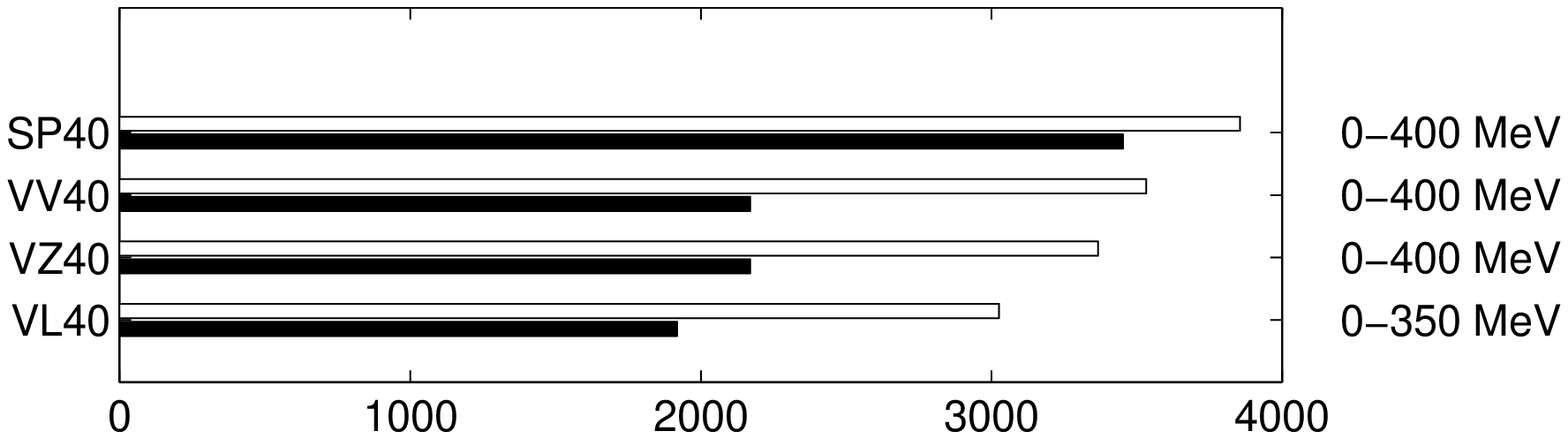,scale=.4}
\caption{The same as shown in Fig. 
\ref{arnsol1} but restricted to sub-threshold data}
\label{arnsol2}
\end{figure}

\begin{figure}[hb]
\centering
\epsfig{file=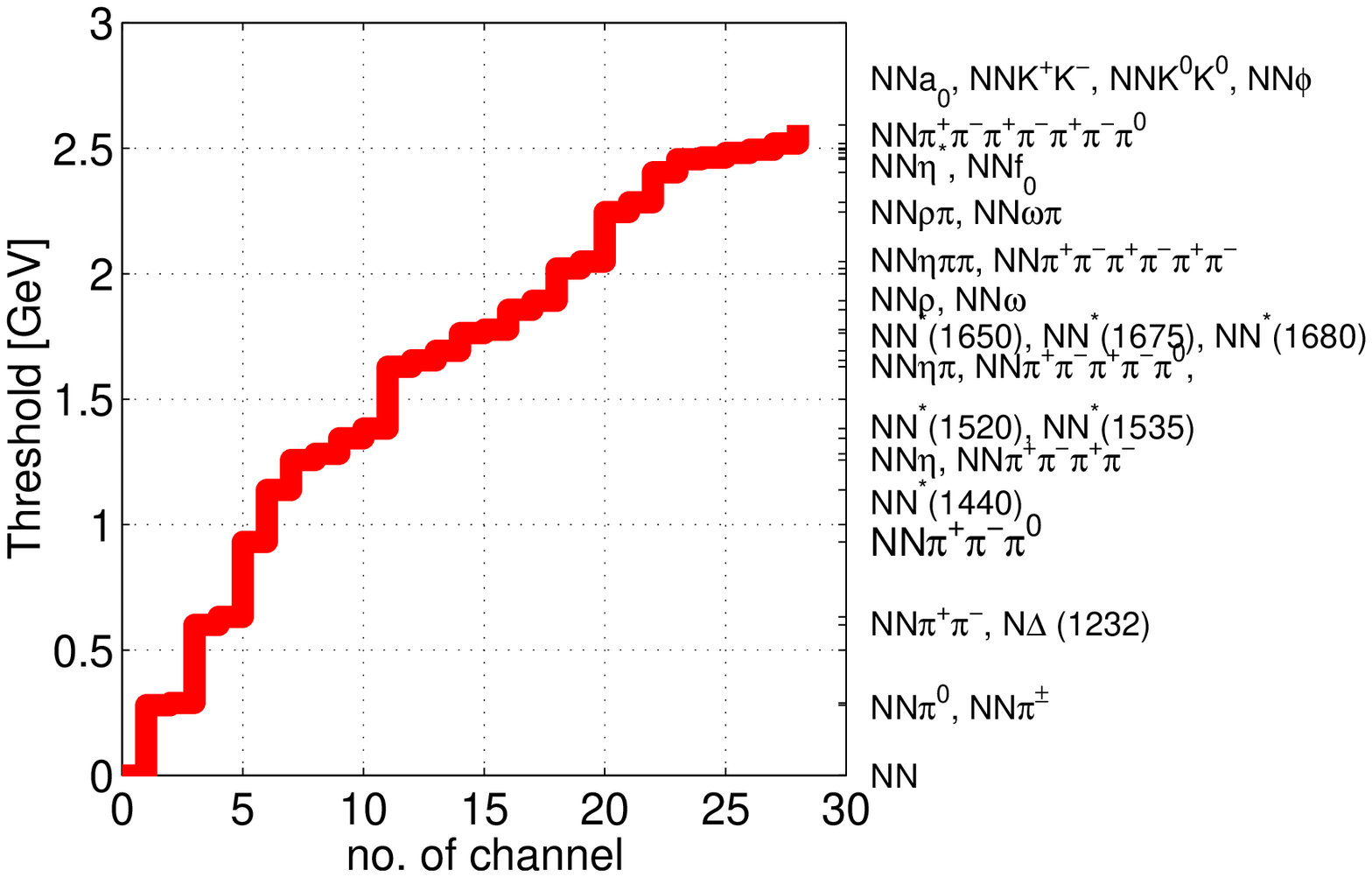,scale=.4}
\caption{NN induced reaction channels}
\label{thresh}
\end{figure}

In view of an optical model study it is timely to review the current
situation of nucleon-nucleon phase shift data below 3 GeV \cite{SAID}. 
NN scattering is a long standing problem for more than five decades 
and has been reviewed under many different aspects, this in progress 
as the data base developed, see Fig. \ref{arnsol1}.
We use as derived quantities the
{\em phase shifts}, taken exclusively from SAID\cite{SAID}. 
The low energy data, $<400$ MeV,  received
particular attention by Arndt {\em et al.}, see solutions in Fig. \ref{arnsol2},
and the Nijmegen group in their NN elastic scattering analysis below $T_\ell\leq350$
MeV\cite{NIJM}. Above pion threshold, $\sim 280$ MeV, reaction channels open in a way shown
in Fig. \ref{thresh}. 
The low energy data, $< 300$ MeV, are subject of NN potentials which we use as reference 
potentials. The classical meson exchange potentials from  Nijmegen, Paris, Bonn,
and our recent OSBEP \cite{Jae98} potential fit the on-shell data within a few percent.
They differ little in their near off-shell t-matrices. 
The Gel'fand-Levitan-Marchenko integral equations relate directly phase shifts 
with off-shell t-matrices.  For convenience this link is  parameterized in 
terms of inversion potentials and the Lippmann-Schwinger equation\cite{Kir91,San96}.
\begin{figure}[ht]
\centering
\epsfig{file=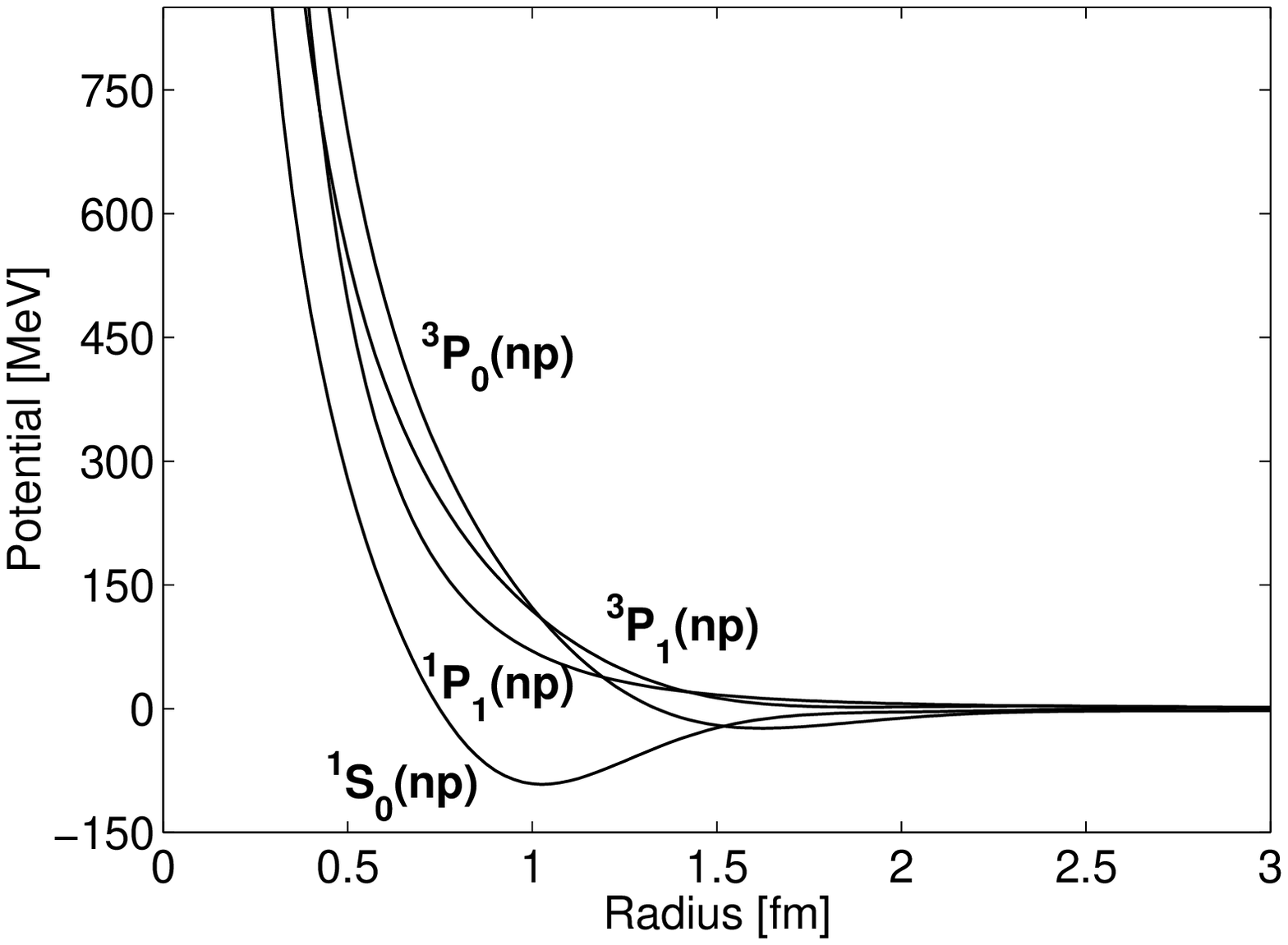,scale=.36}
\epsfig{file=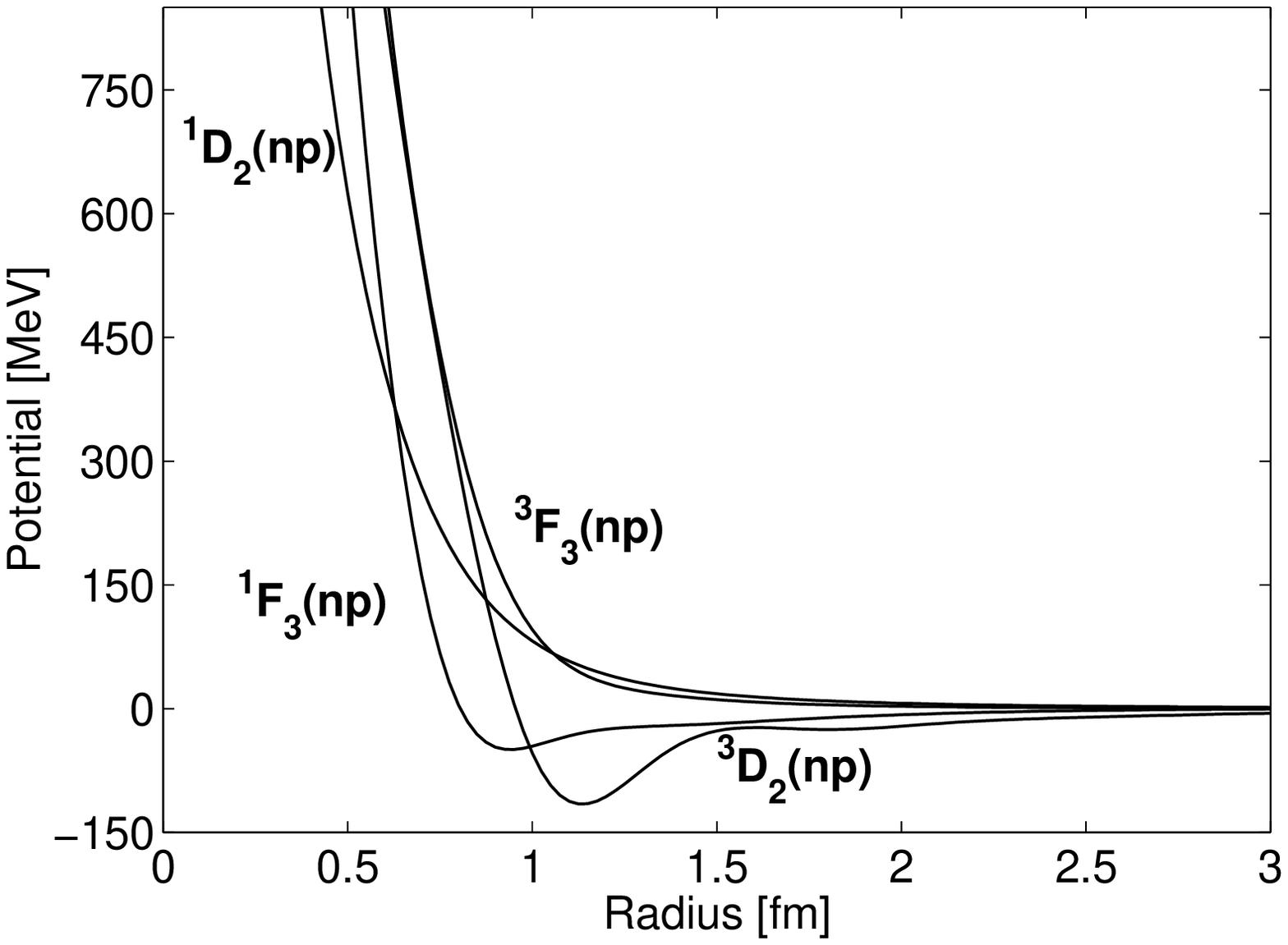,scale=.36}
\caption{Nucleon-nucleon inversion potentials using SP00 phases}
\label{ivspot}
\end{figure}
In Fig. \ref{ivspot} is shown a sample of potentials which reproduce the low 
energy phase shifts. They allow  to distinguish a long range Yukawa tail, 
a medium range attraction $\sim 1-2$ fm and
a short range repulsion. From these potentials, equally valid for the boson exchange
potentials, we draw the conclusion that the long and medium range NN interaction diminish
in their
importance for energies  above 500 MeV. For projectiles with $T_\ell>1$ GeV
the repulsive core is  remaining and predominantly causes scattering. The core
strength reaches a value of 1 GeV at $\sim 0.5$ fm which may be viewed as 
the classical turning point.

The real new feature above 300 MeV is the occurrence of inelasticity and thus breakdown
of unitarity of the elastic channel S-matrix. However, the above mentioned
potentials may still
be compared with the real parts of phase shifts.
From the theoretical contents of boson exchange models and their similar parameterization
it is expected that they extrapolate from the fitted domain $<300$ MeV towards higher
energies with the same closeness as in the fitted domain.
As is shown
in Figs. \ref{sglphasen} and \ref{cplphasen}
 this is not the case. We argue that off-shell differences
 in the high quality potentials are
essentially caused by these high energy differences, made visible in the on-shell data
$0.3-3$ GeV.

The actual phase shifts cover the energy range up to 3
GeV for pp scattering and 1.2 GeV for np scattering. This is ten times more than 
what the classical potentials
describe. Their extrapolation towards higher energies is quite different
for each solution and by dint of its construction only inversion potentials follow closely
the real part of the phase shifts at all energies.

\section{Optical Model Analysis}

\begin{figure}[H]
\centering
\epsfig{file=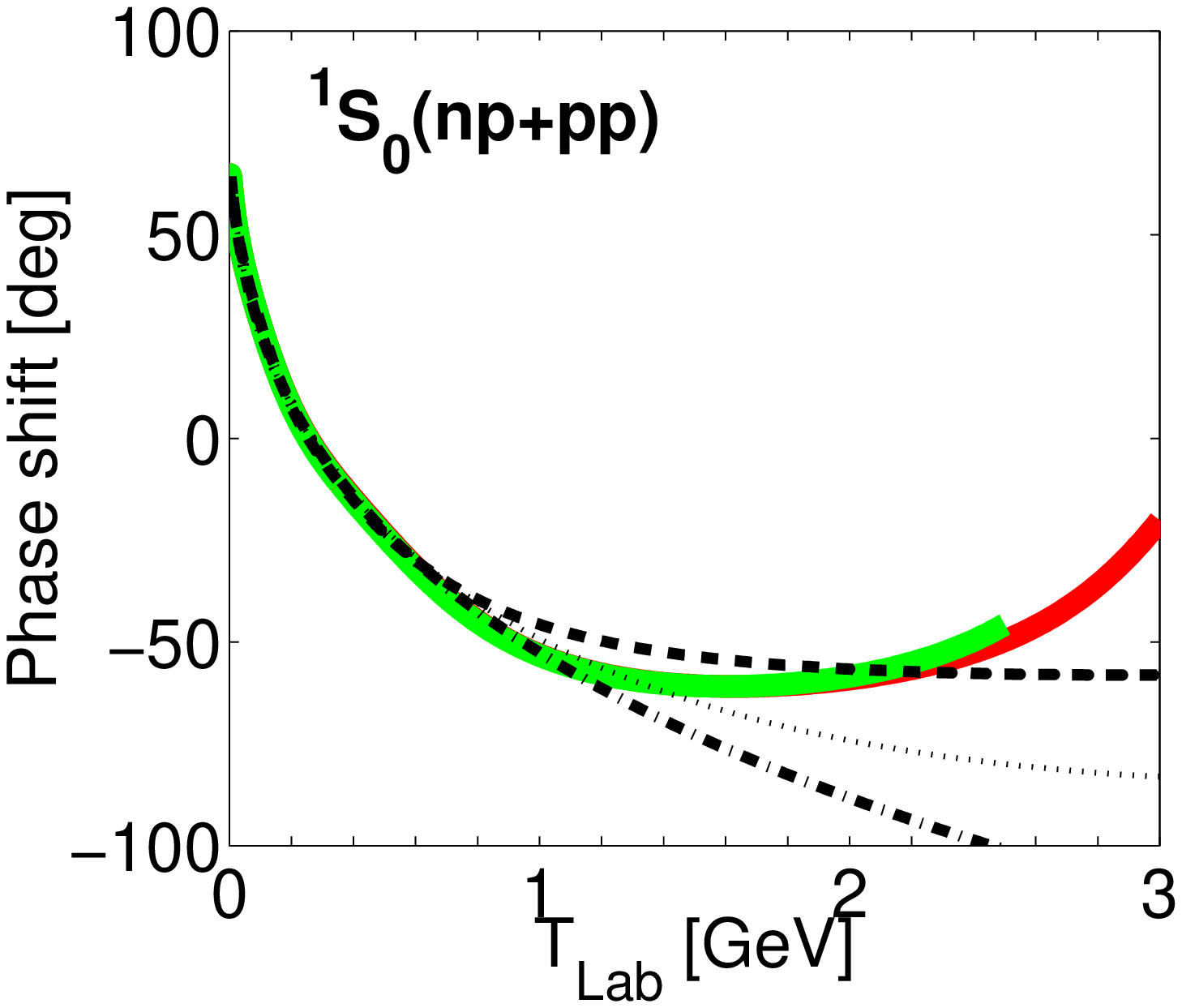,scale=.3}
\epsfig{file=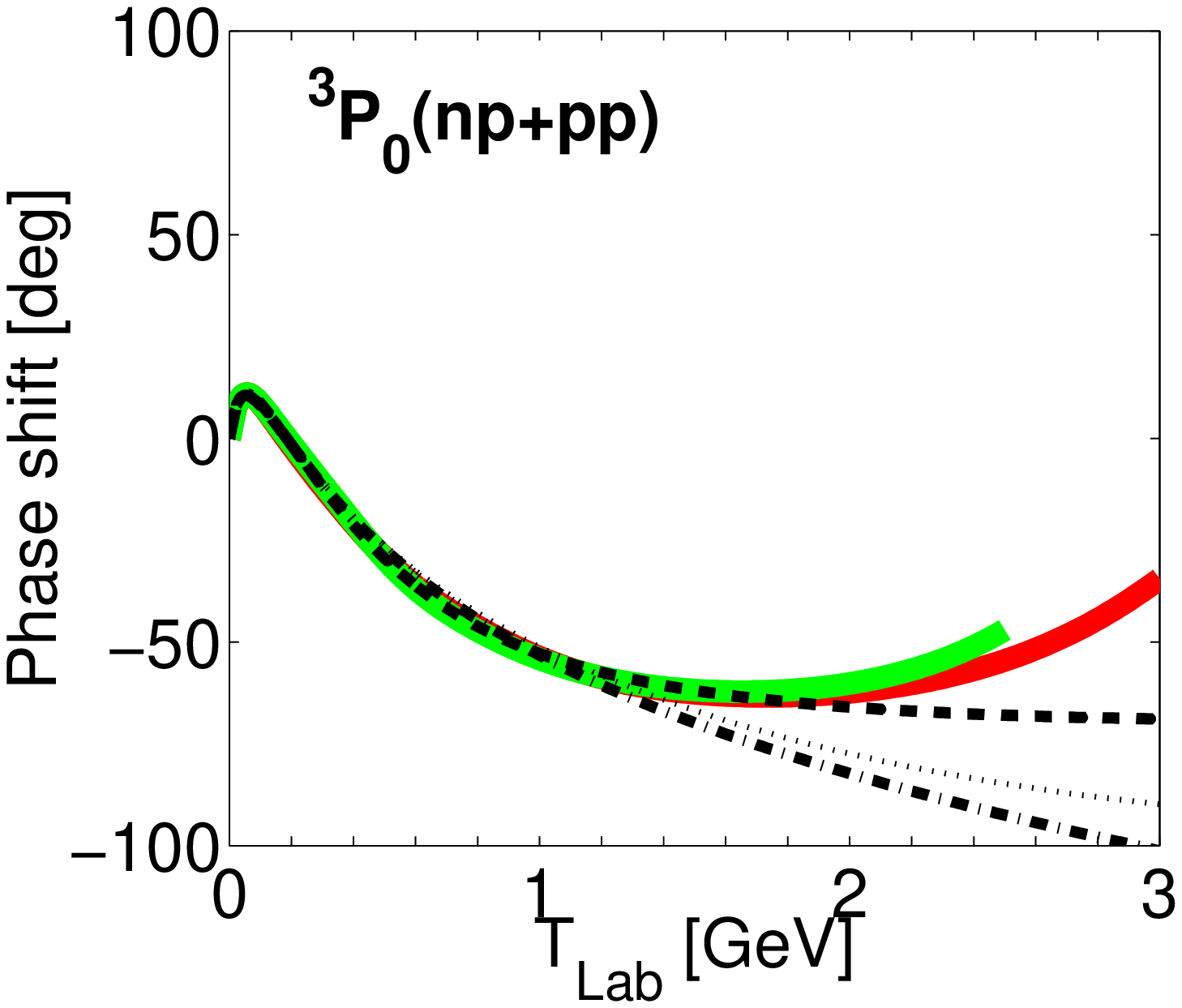,scale=.3}
\\
\epsfig{file=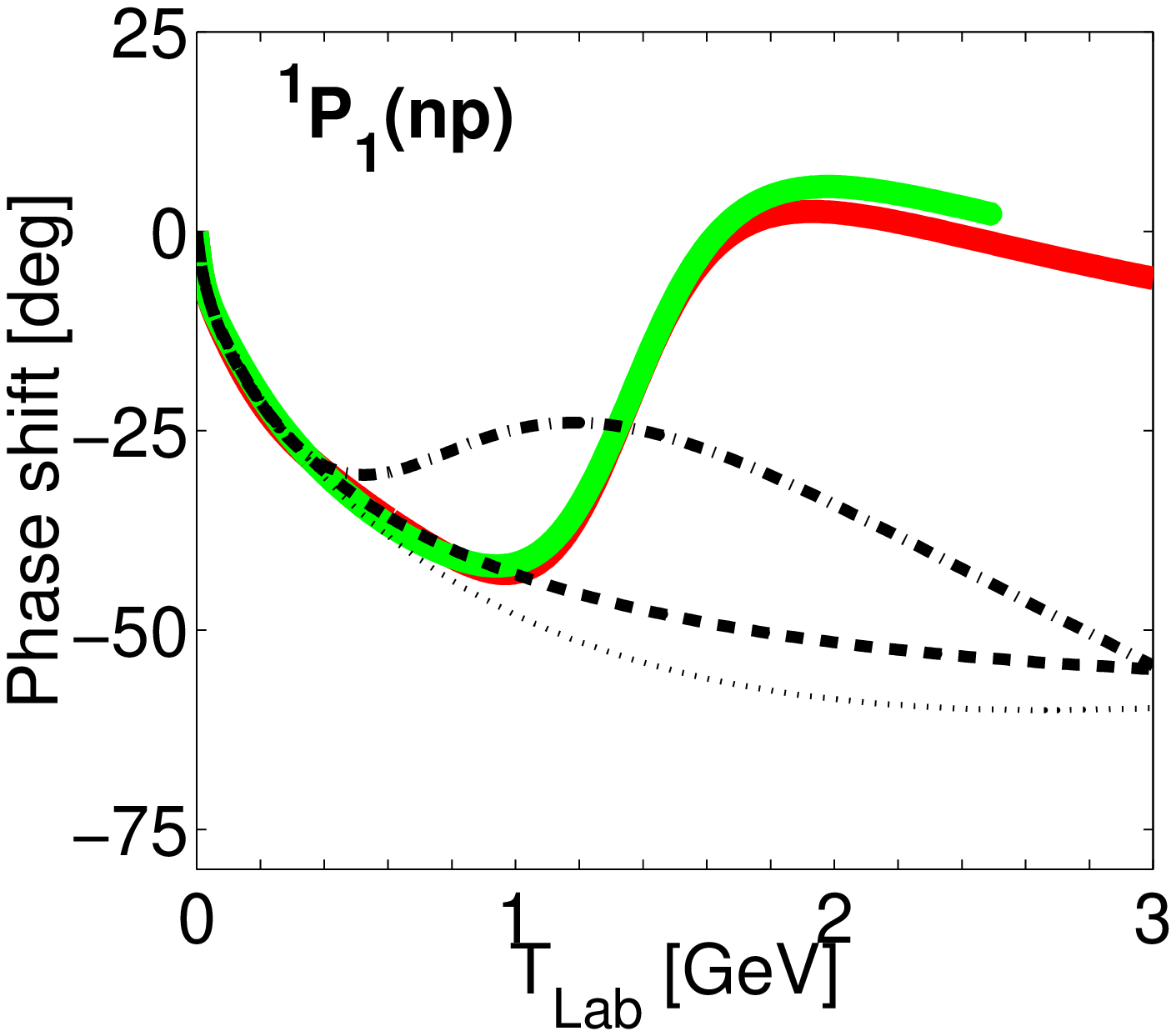,scale=.3}
\epsfig{file=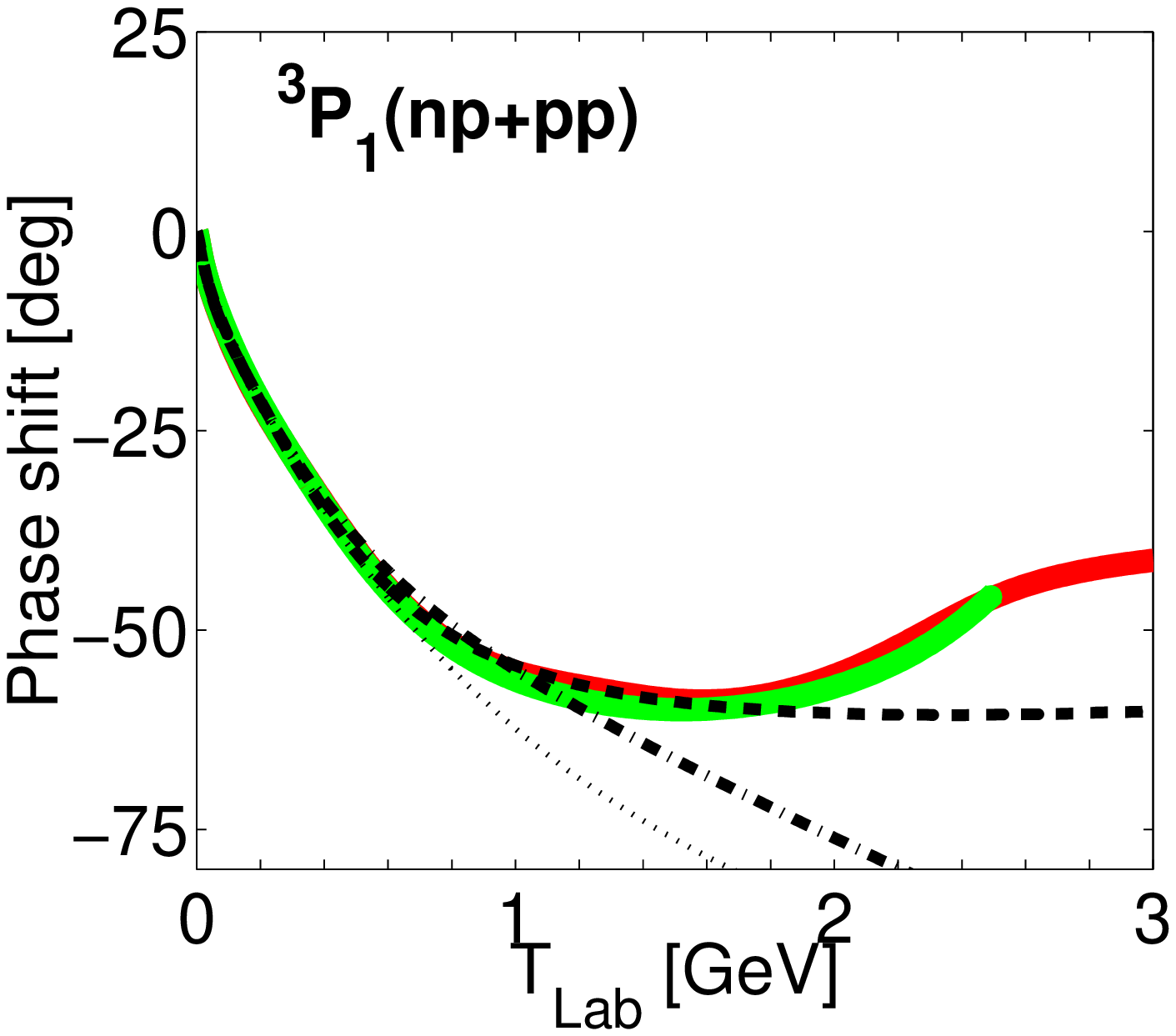,scale=.3}
\\
\epsfig{file=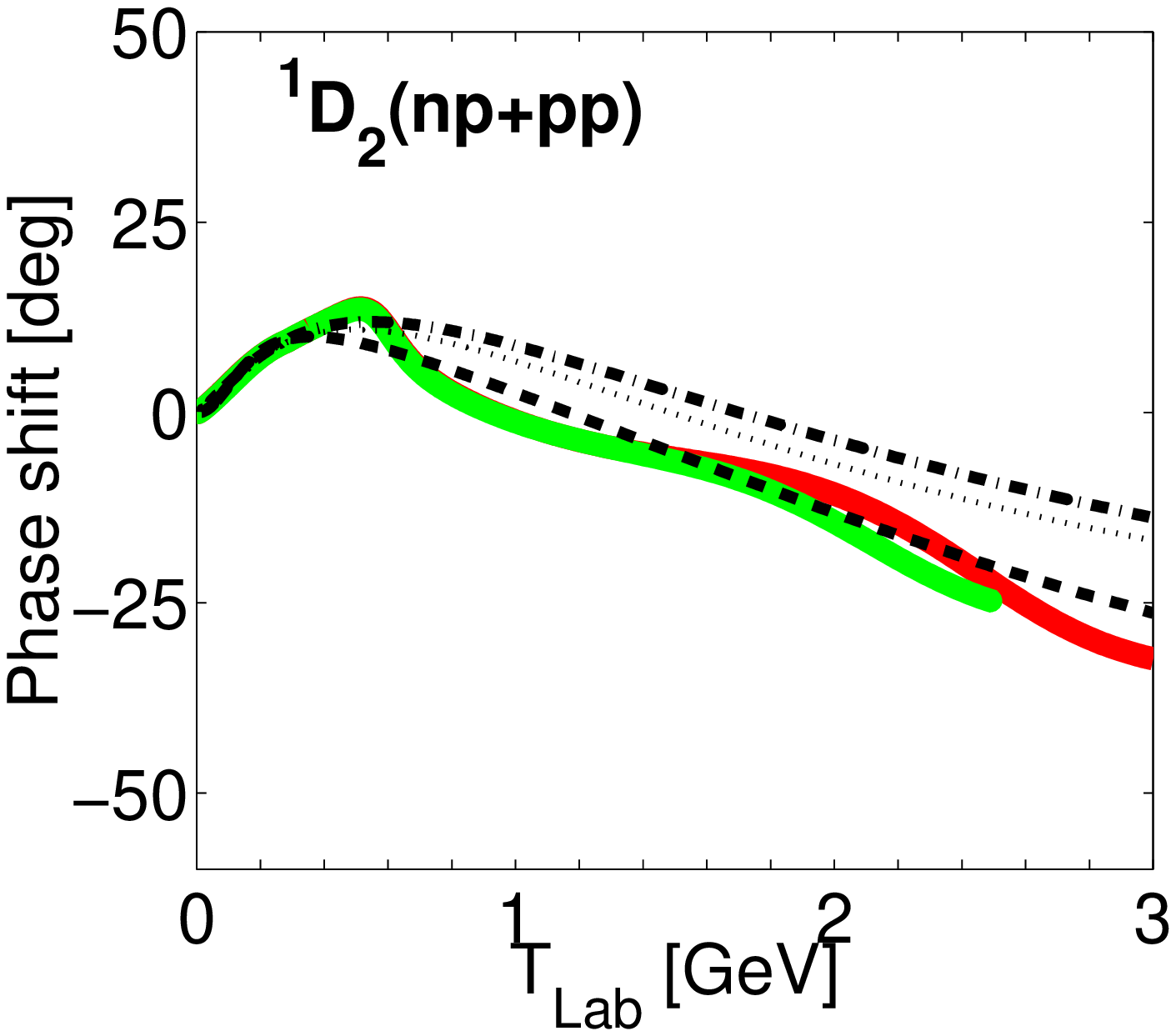,scale=.3}
\epsfig{file=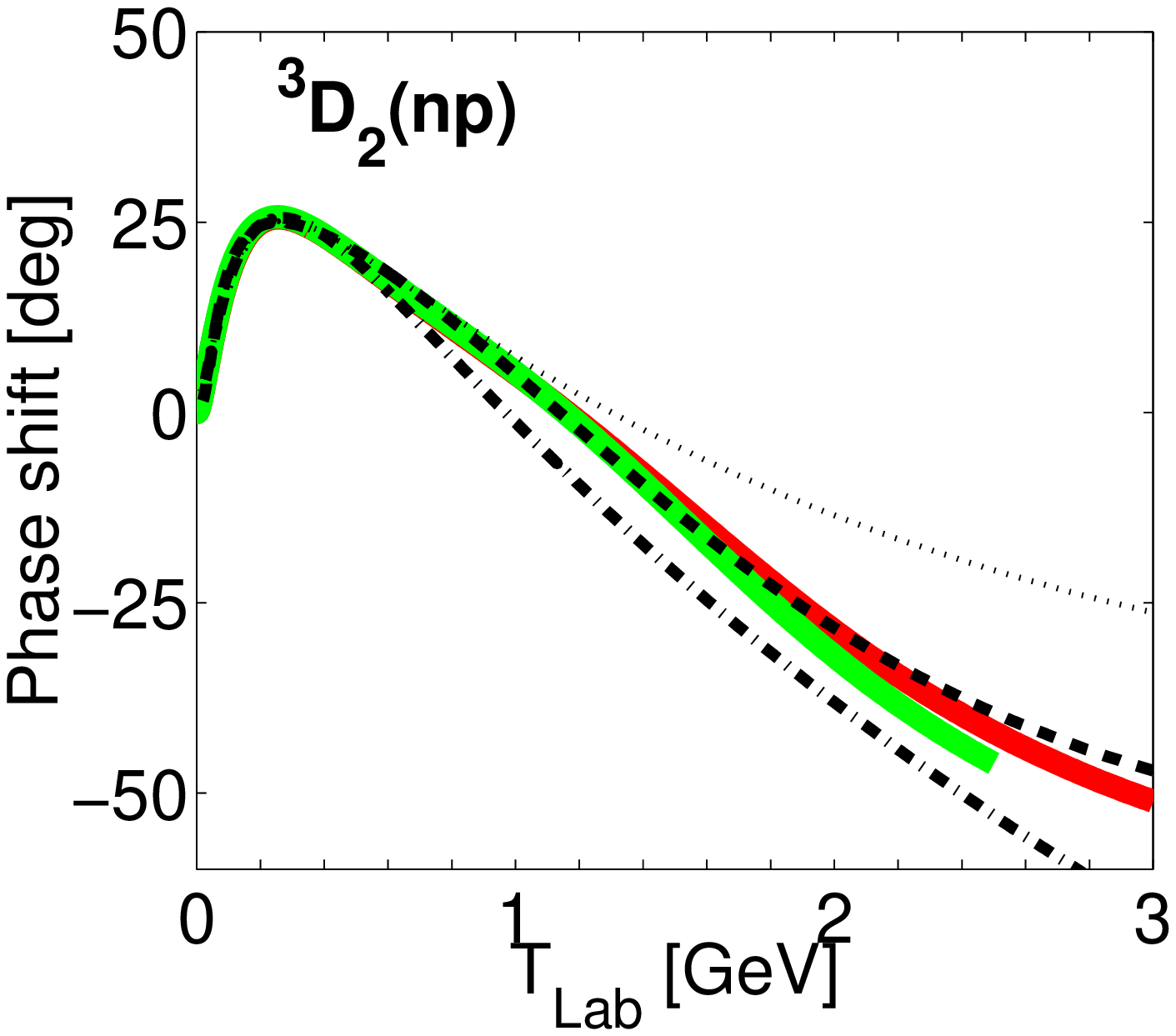,scale=.3}
\\
\epsfig{file=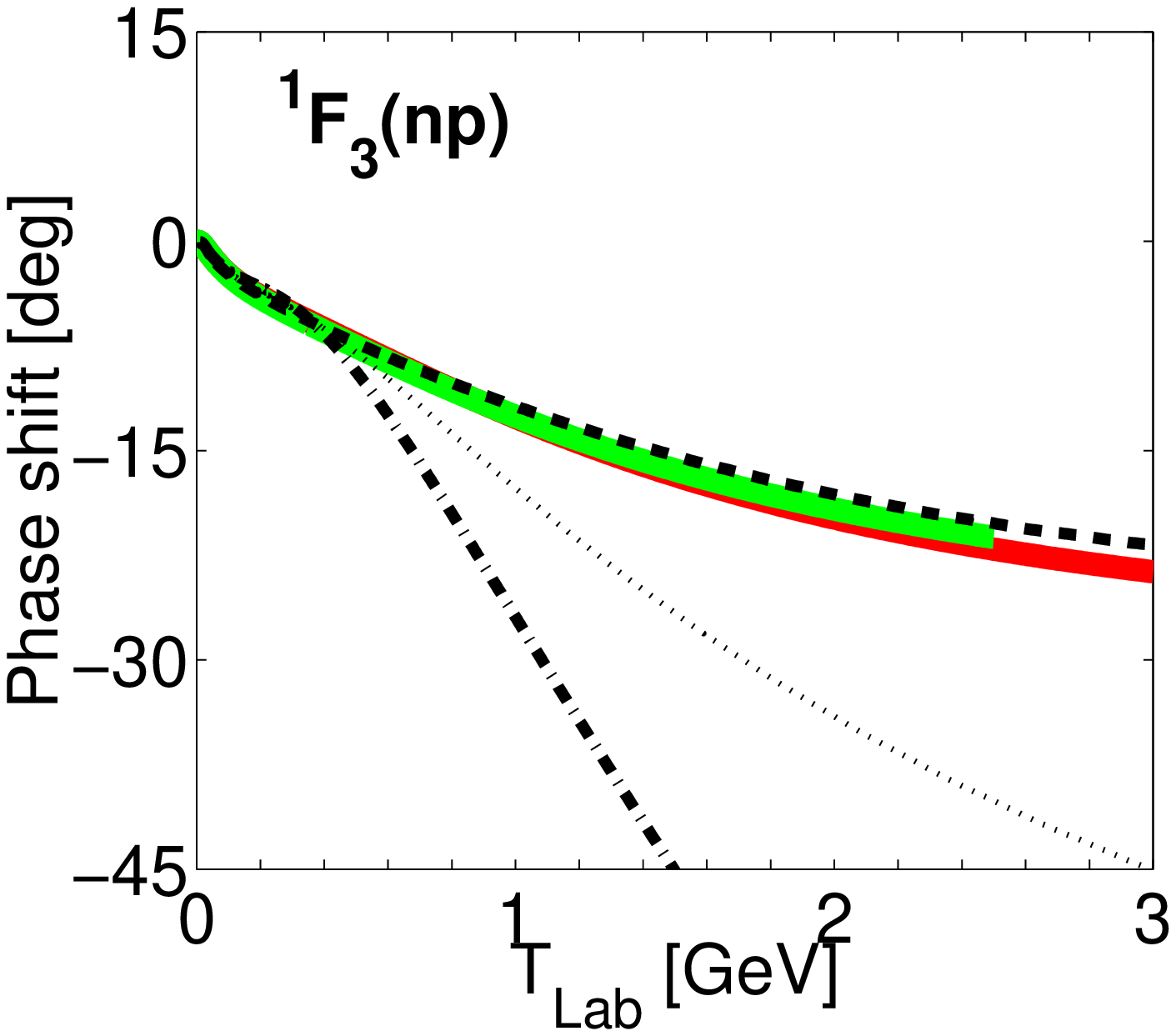,scale=.3}
\epsfig{file=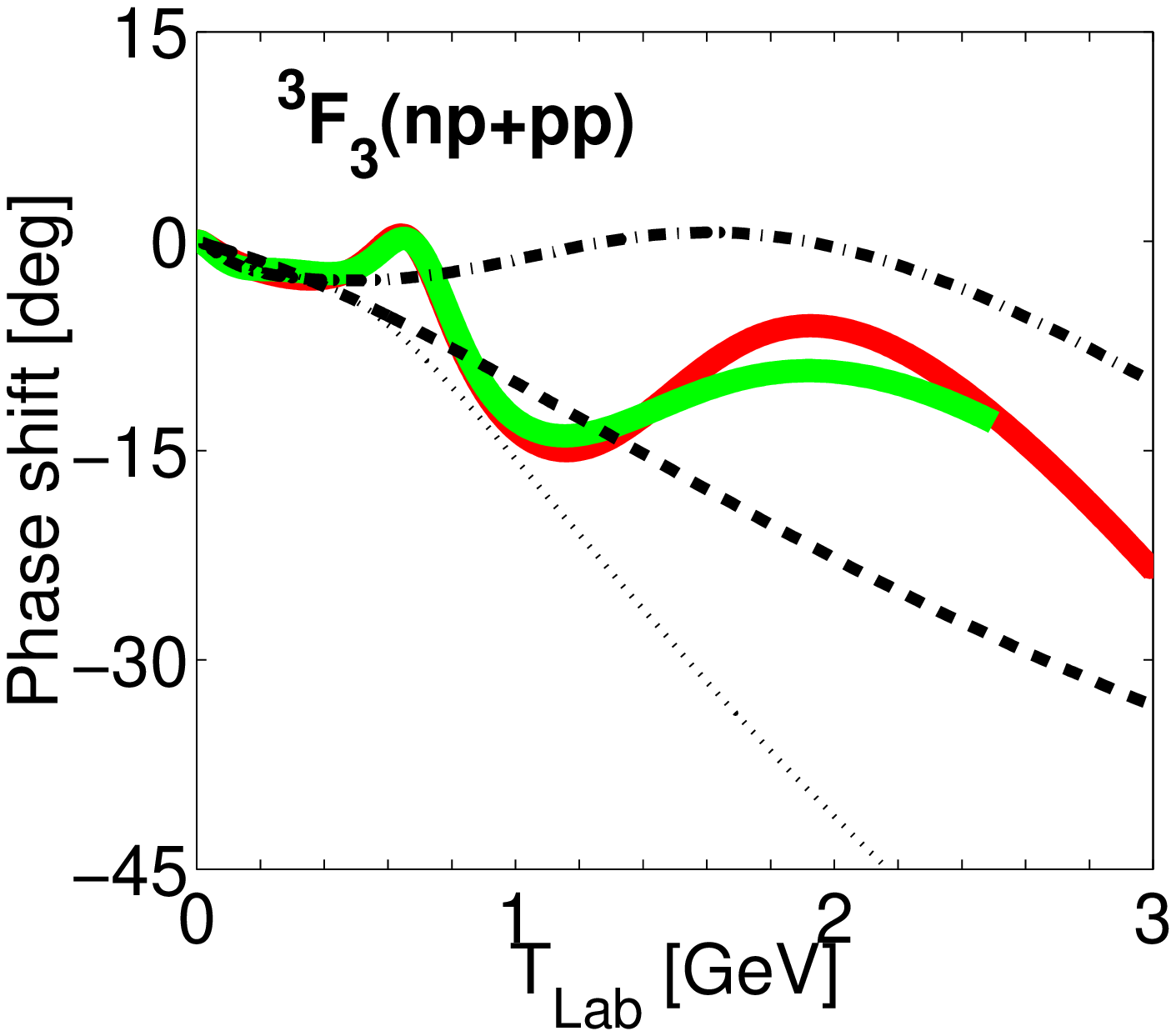,scale=.3}
\caption{SM97 (shaded, $<2.5$ GeV) and FA00 (shaded, $<3$ GeV)
 phase shifts and reference phase 
shifts from HH-inv (dashed),Nij-1 (dash-dotted) and AV18 (dotted)}
\label{sglphasen}
\end{figure}

\begin{figure}[htb]
\centering
\epsfig{file=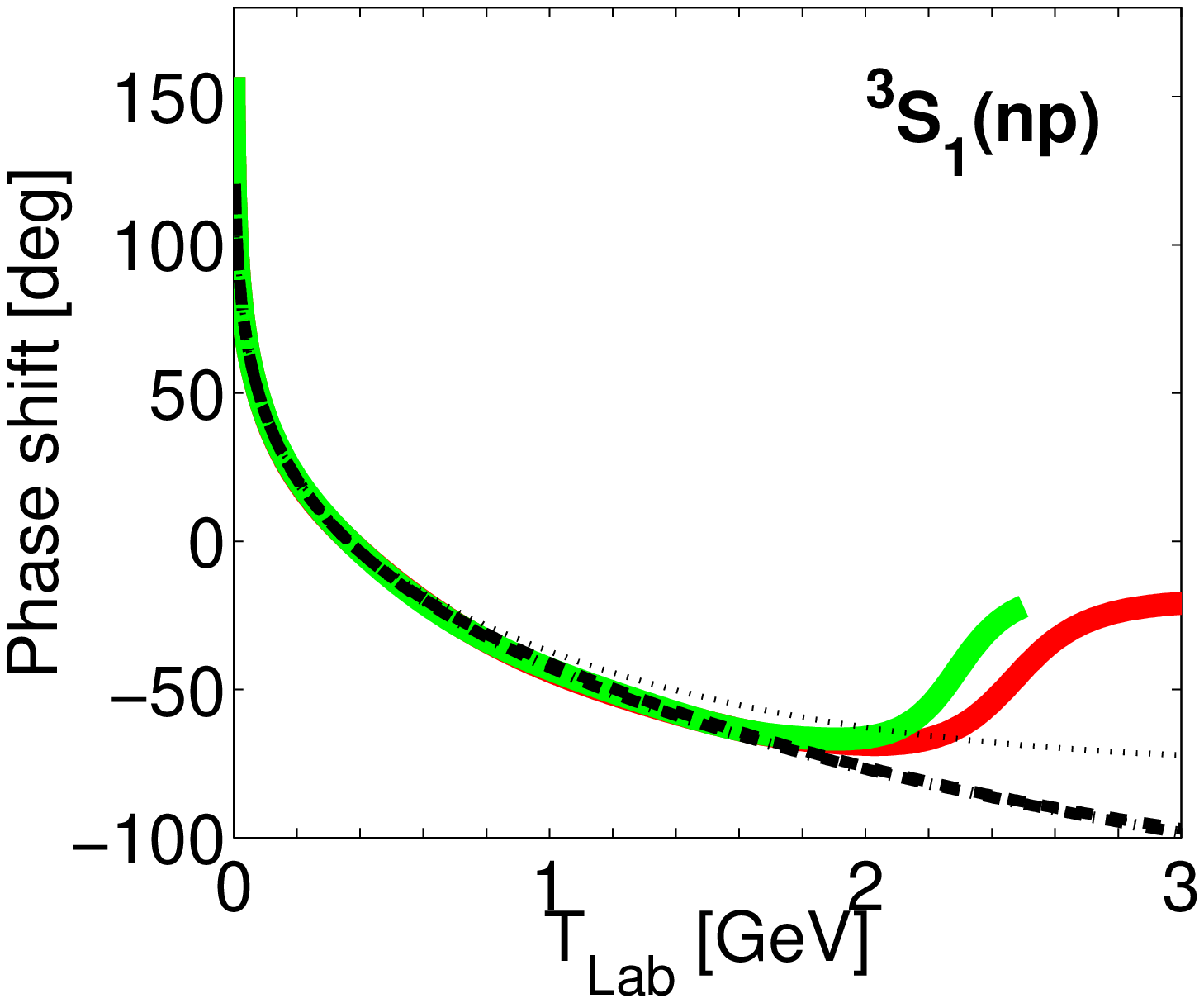,scale=.3}
\epsfig{file=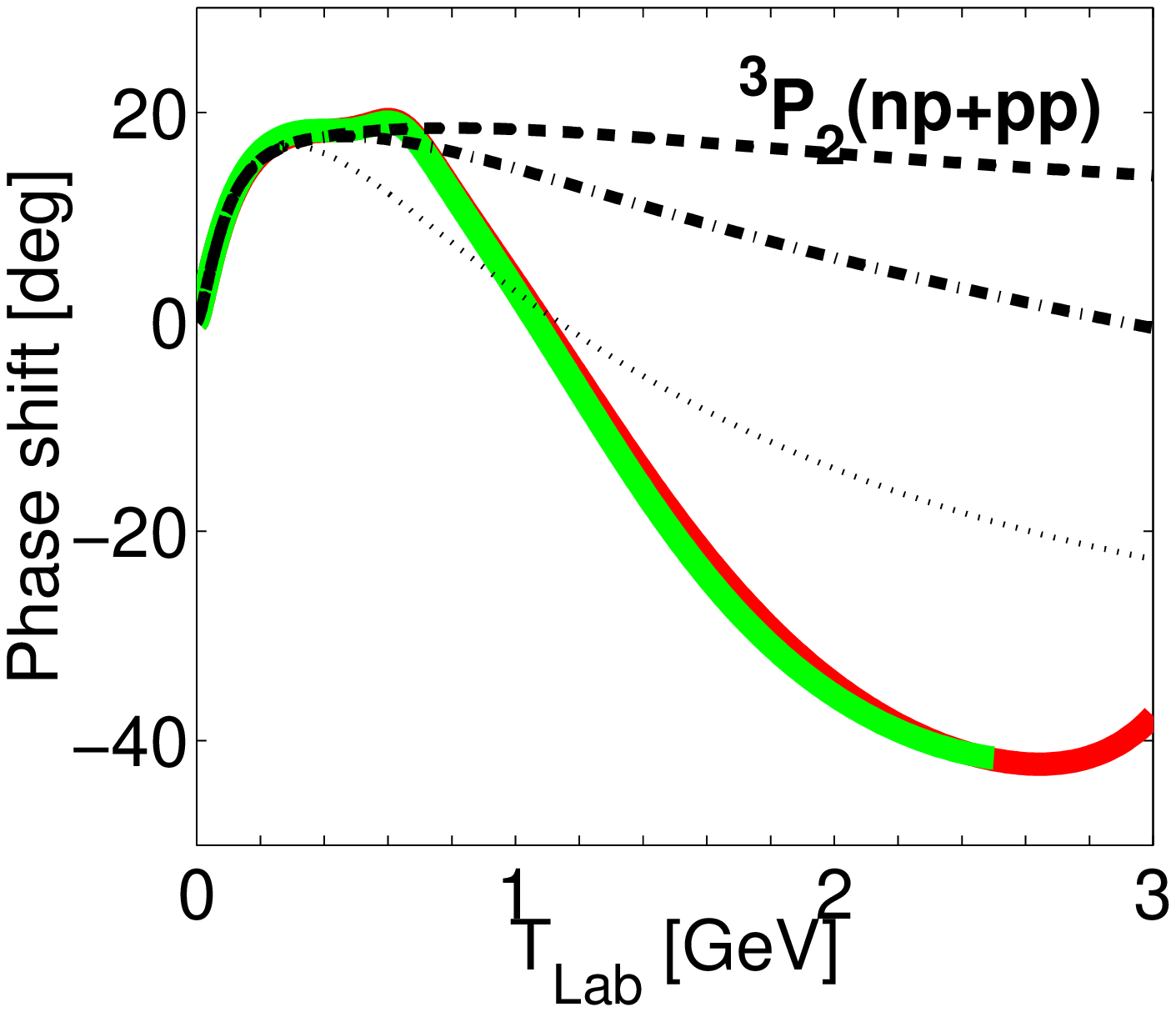,scale=.3}
\\
\epsfig{file=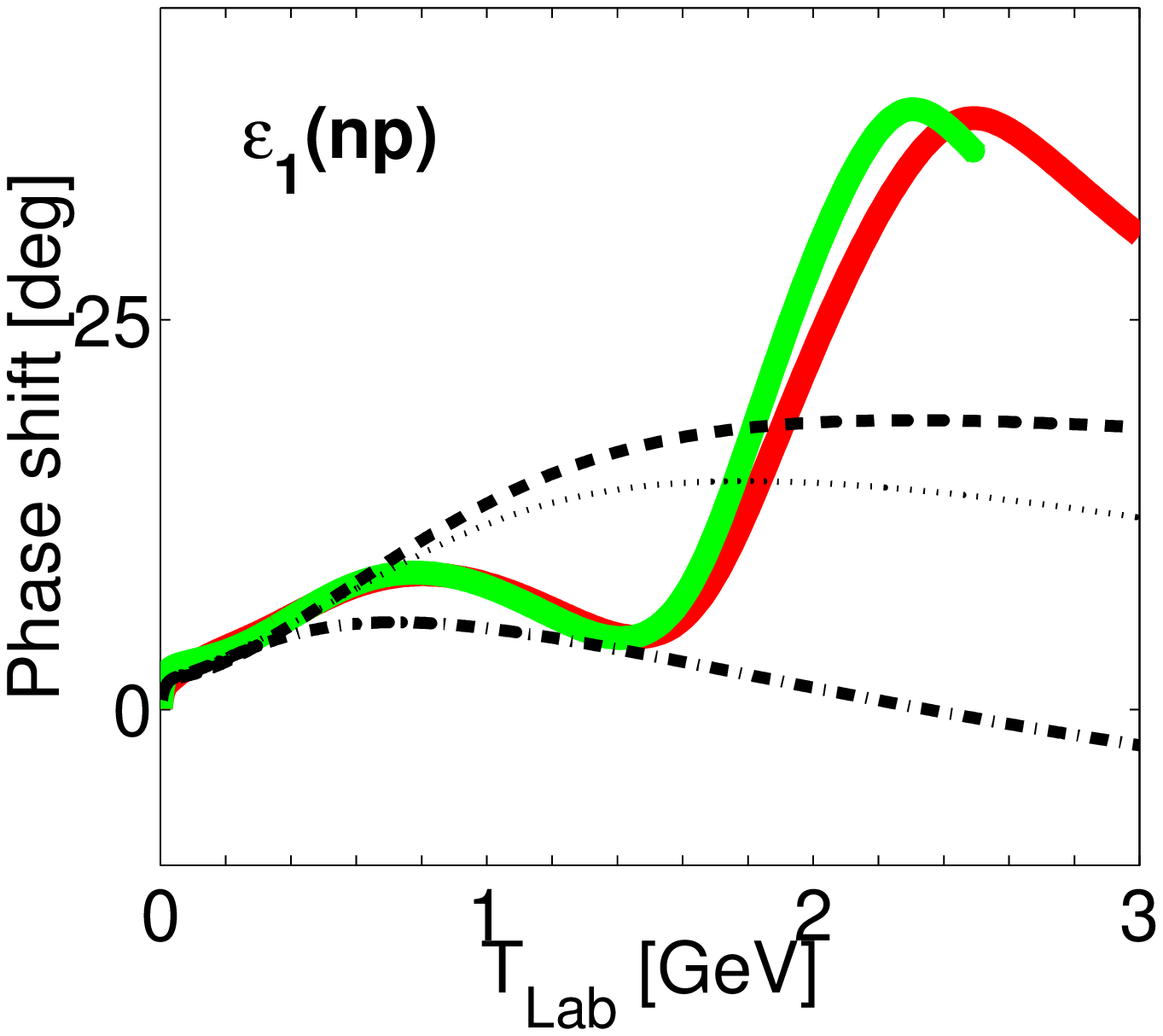,scale=.3}
\epsfig{file=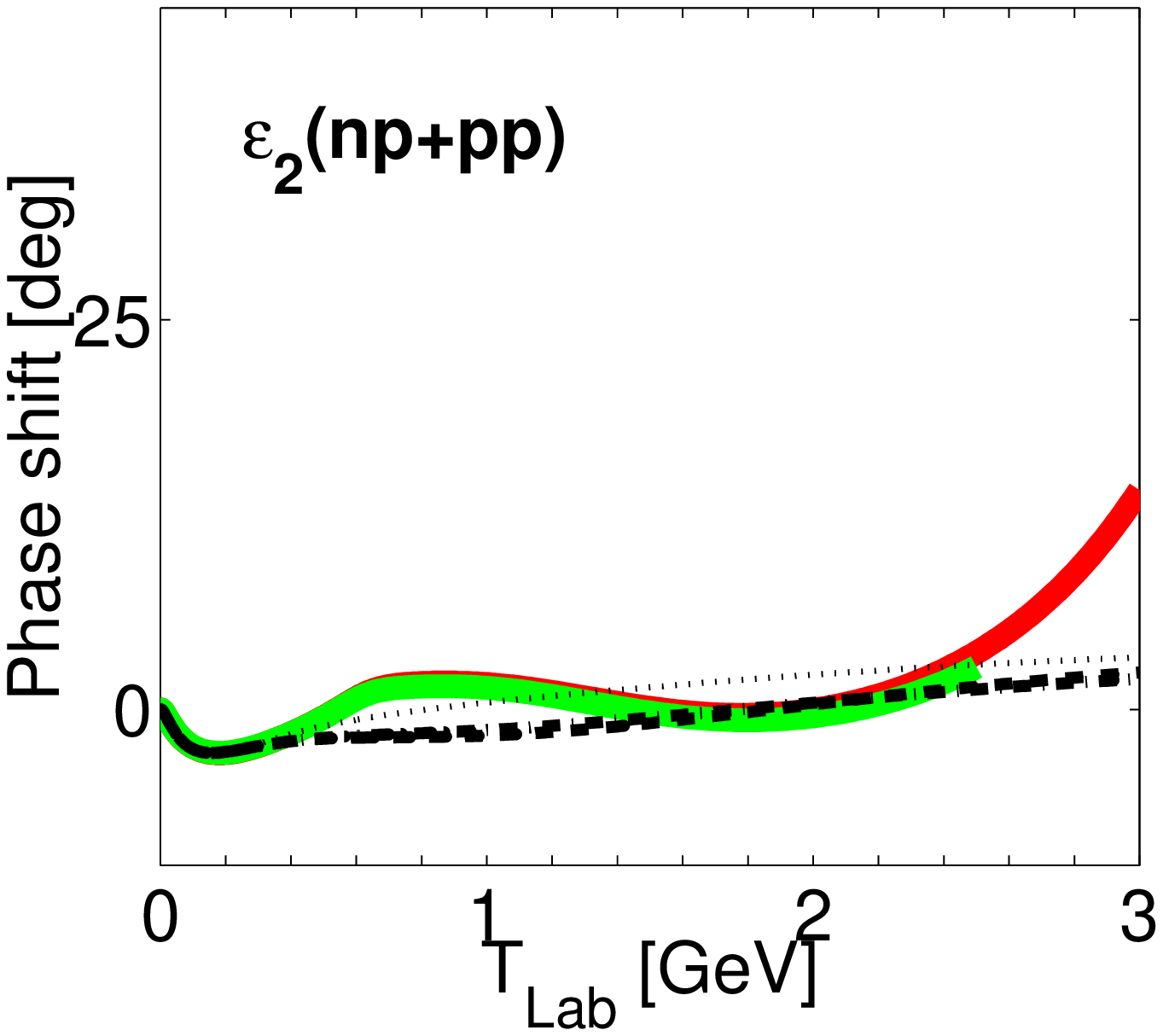,scale=.3}
\\
\epsfig{file=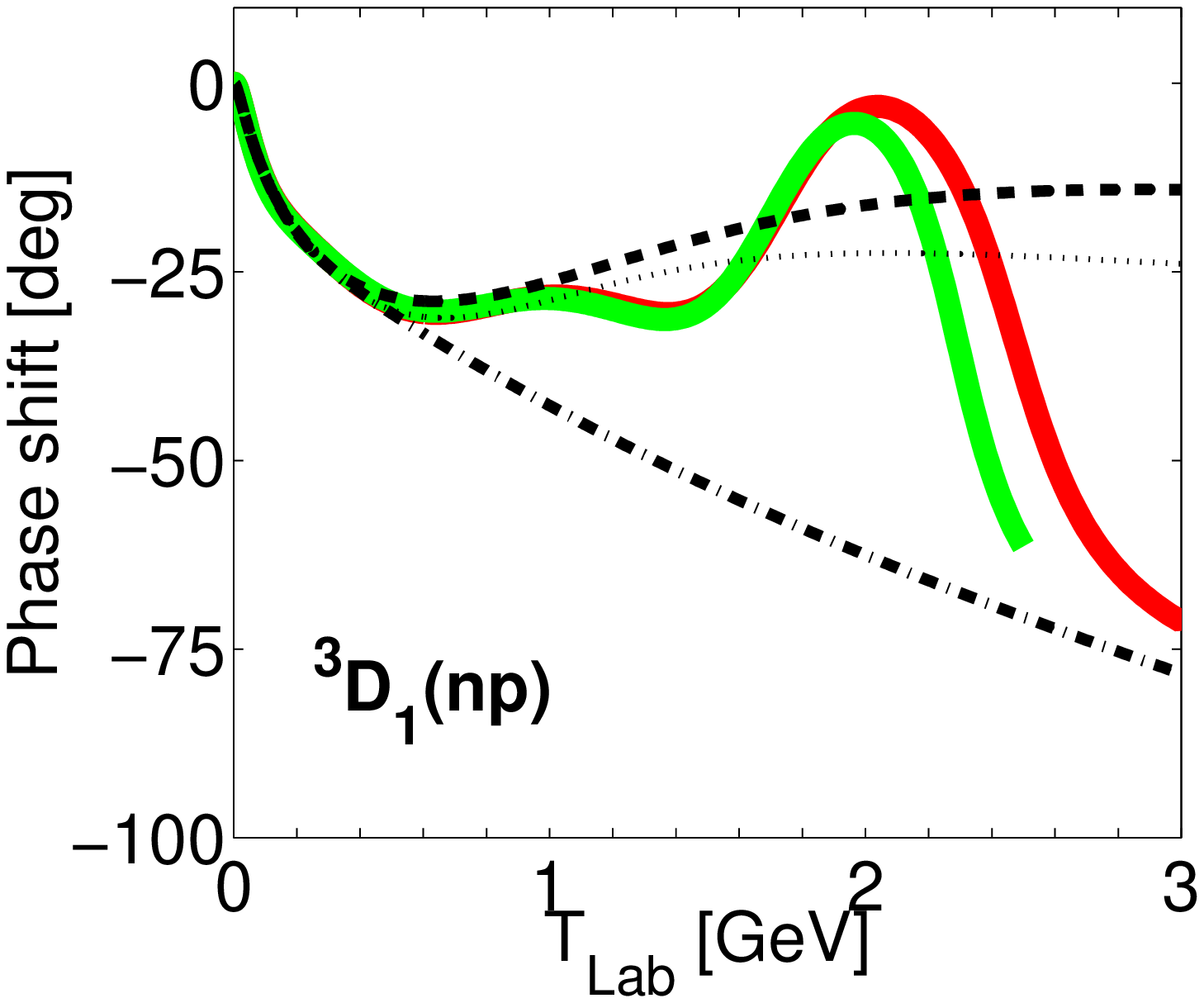,scale=.3}
\epsfig{file=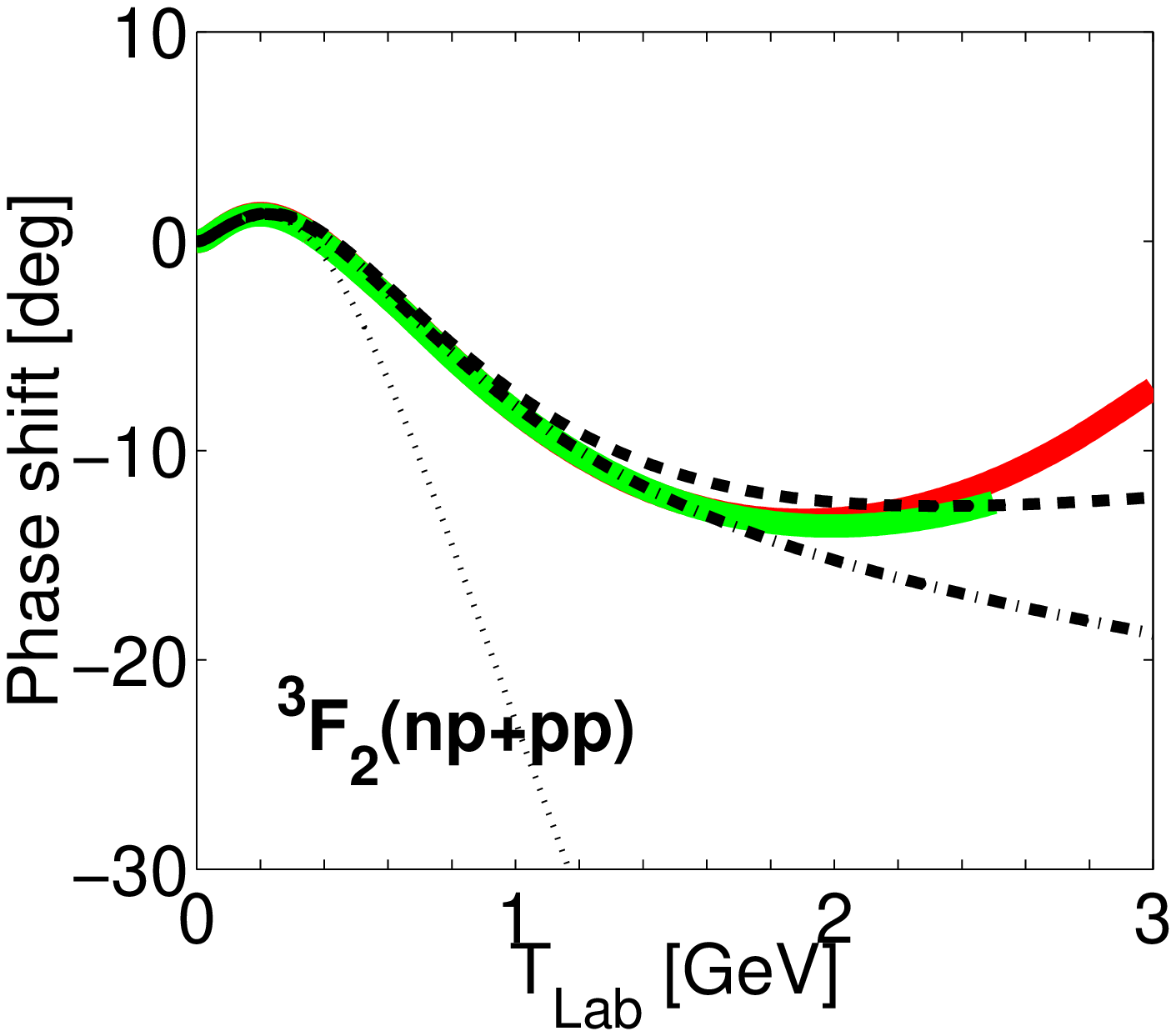,scale=.3}
\caption{As previous figure but for the coupled channels 
$^3SD_1$ and $^3PF_2$}
\label{cplphasen}
\end{figure}

The loss of unitarity
is met by an optical model potential. Such potential approaches are well
known from nucleon-nucleus scattering and are used for NN scattering as 
well\cite{Neu91,Ger97,Fal99}. In another contribution to these proceedings, where  
the theoretical and technical background of the optical model was introduced, 
we distinguish real reference potentials
and complex valued separable potentials (or complex boundary conditions instead).
This model is used together with an appraisal of the data  shown next.
In Figs. \ref{sglphasen} and \ref{cplphasen}
 the SM97 and FA00 solutions are shown together with potential model phase
shifts, $J\leq 3$, Nij-1, AV18, HH-inv. Arndt's solutions SM97 and FA00 are stable. 
Further on, we shall use the solution FA00 (it differs from SP00 and SM00 only marginally).
In Fig. \ref{ompresults} we show the results of optical model calculations
in which the reference potentials are Gel'fand-Levitan-Marchenko potentials
with a rank one separable potential (harmonic oscillators
 $\phi_\ell(r,\hbar\omega=450\mbox{MeV})$). The reference potential phase shifts
are shown together (as dotted lines) with the real phase shift data (continuous 
solution) and radial distributions of probabilities and loss of flux. 
\begin{equation} 
\rho(r,\alpha,E)  = Trace\ \Psi^\dagger (r,\alpha,E)\Psi(r,\alpha,E)r^{-2}
\end{equation}
and the loss of flux, from the continuity equation
\begin{equation}
{\partial\over \partial t}\rho(r)+(\vec \nabla\cdot\vec j)=0,
\end{equation}
\begin{equation}
(\vec \nabla\cdot\vec j) : = Trace\ {i\over \hbar}
<\Psi(\alpha,E)|{\cal V}(.,r)-{\cal V}^\dagger(r,.)|\Psi(\alpha,E)>.
\end{equation}

A glance upon reaction and inelastic channels is useful.
In Fig. \ref{thresh}  are channels plotted against threshold energy. 
The region  0.5-1 GeV is dominated by the $\Delta$(1232) resonance whereas 
several $N^*$ resonances shape the region  1-2 GeV\cite{NUCDAT}. 
In the high energy region, $\sim 2.5$ GeV, we see many production
channels opening. These reactions couple within several partial waves so the
effects are smoothly distributed and we have good conditions for an analysis with 
a smooth optical potential. 

\begin{figure}[ht]
\centering
\epsfig{file=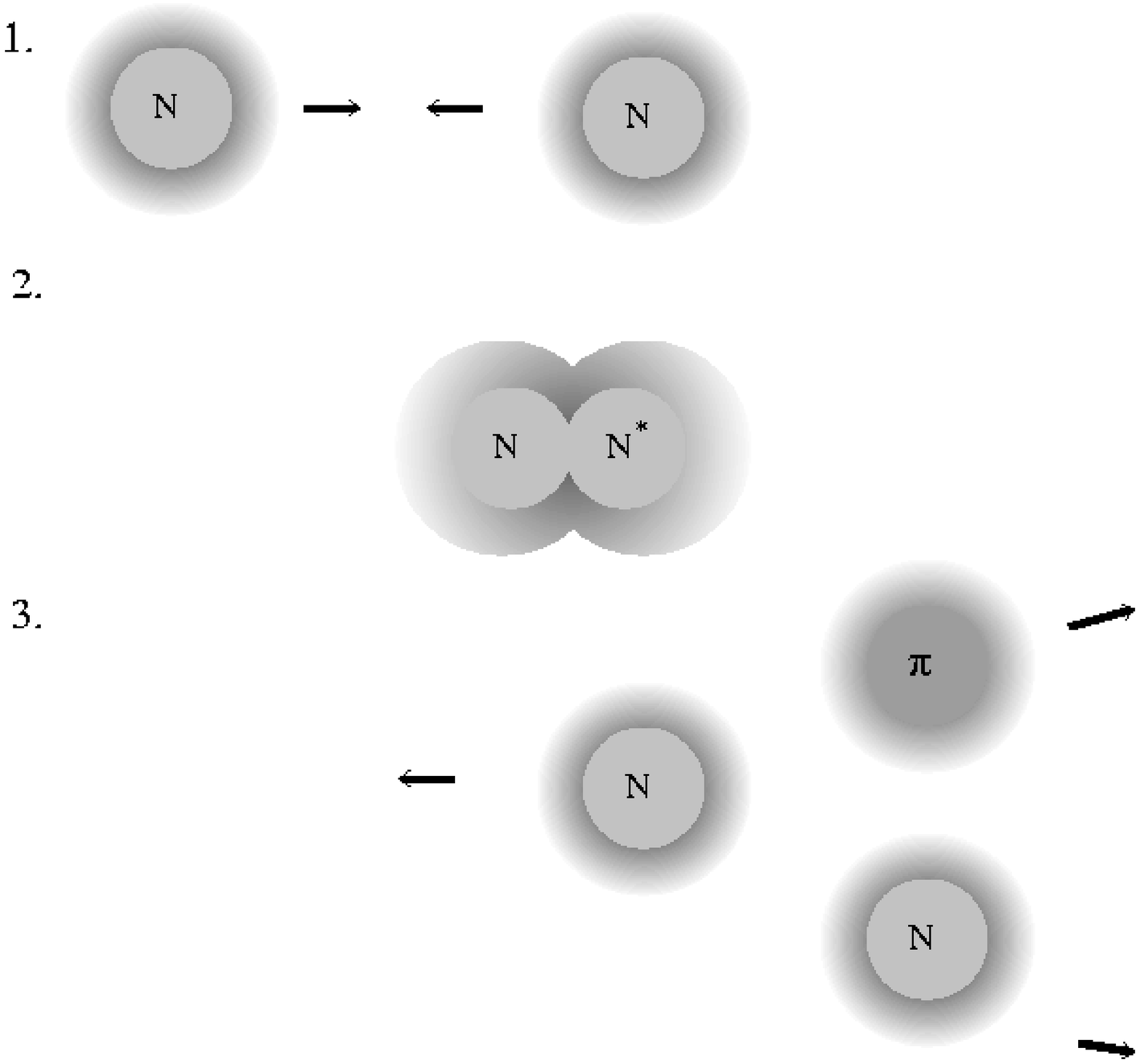,scale=0.35}
\caption{Guide to mental image of NN interaction sequence}
\label{nucball}
\end{figure}

A mental image of the reaction sequence is shown in Fig. \ref{nucball}. 
First, the two nuclei approach each other and their respective  
meson clouds are exerting the boson exchange forces upon each other. 
The nuclei are still well separated (long and medium range forces). Next, after  
further approach, one or both nucleons are intrinsically 
excited into intermediate states which
rapidly de-excite by  particle  emission. The scattering may be either elastic or
 inelastic into three and more body systems.
The image suggests no fusion of the two nucleons, which is a feature quite different 
from  the studied $\pi$N resonance formation. The Pauli principle is very effective
in generating the repulsive core and it appears also  effective in the conservation
of individual nucleons. At $T_\ell= 2-3$ GeV they are well separated in momentum space. 

\section{Probability and Loss of Flux}

\begin{figure}[H]\centering
\epsfig{file=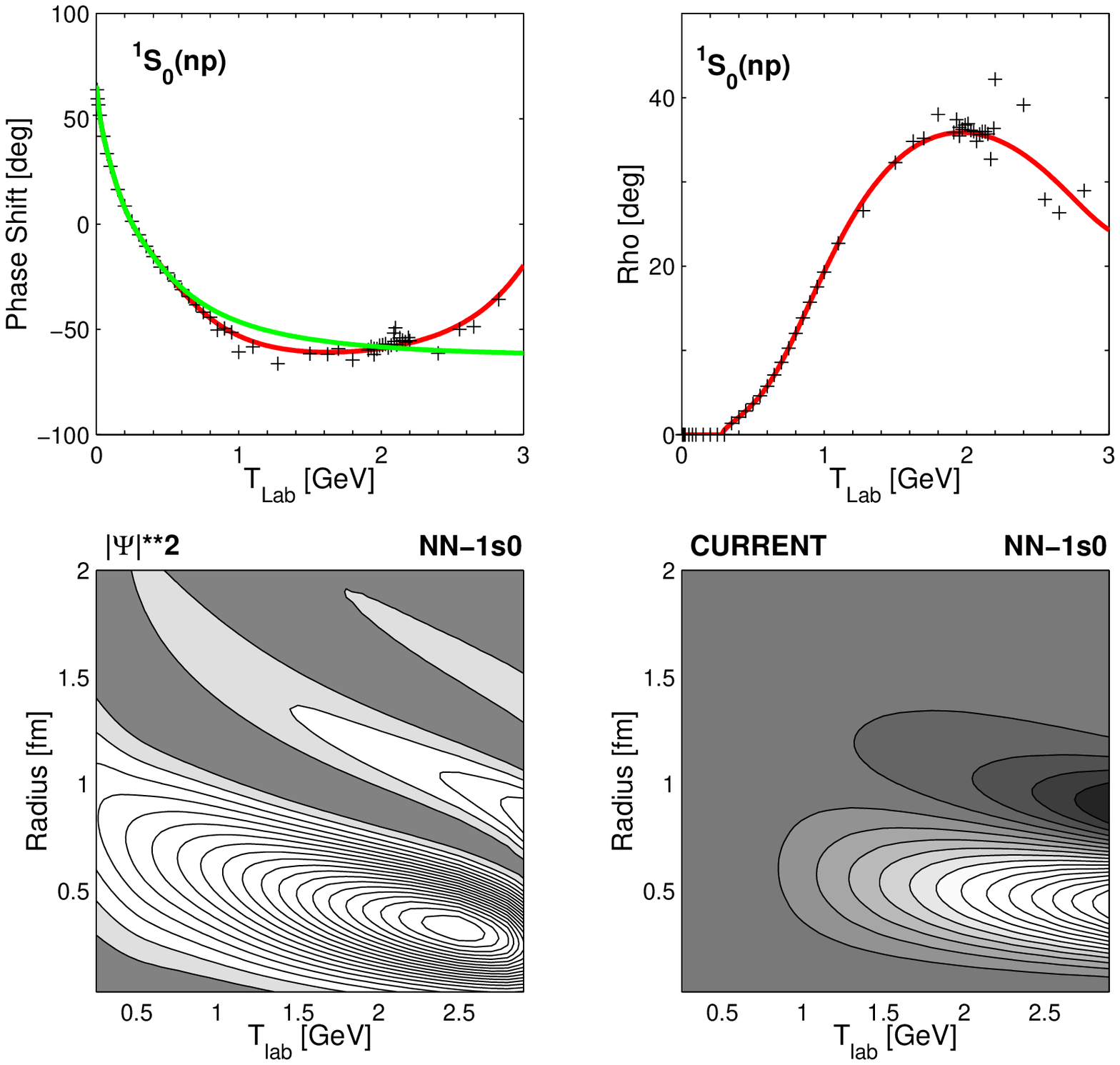,scale=.3}
\epsfig{file=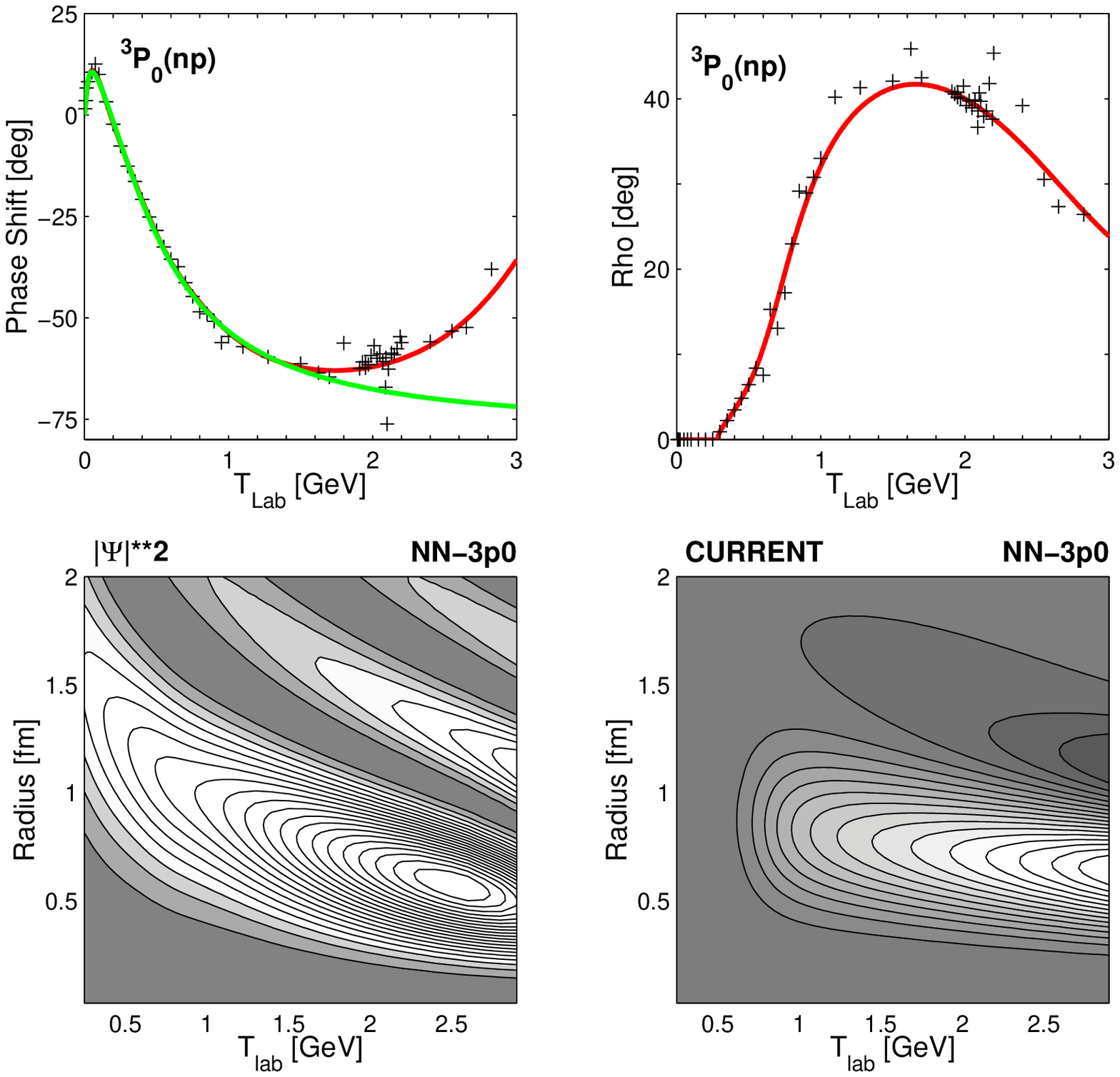,scale=.3}
\\
\epsfig{file=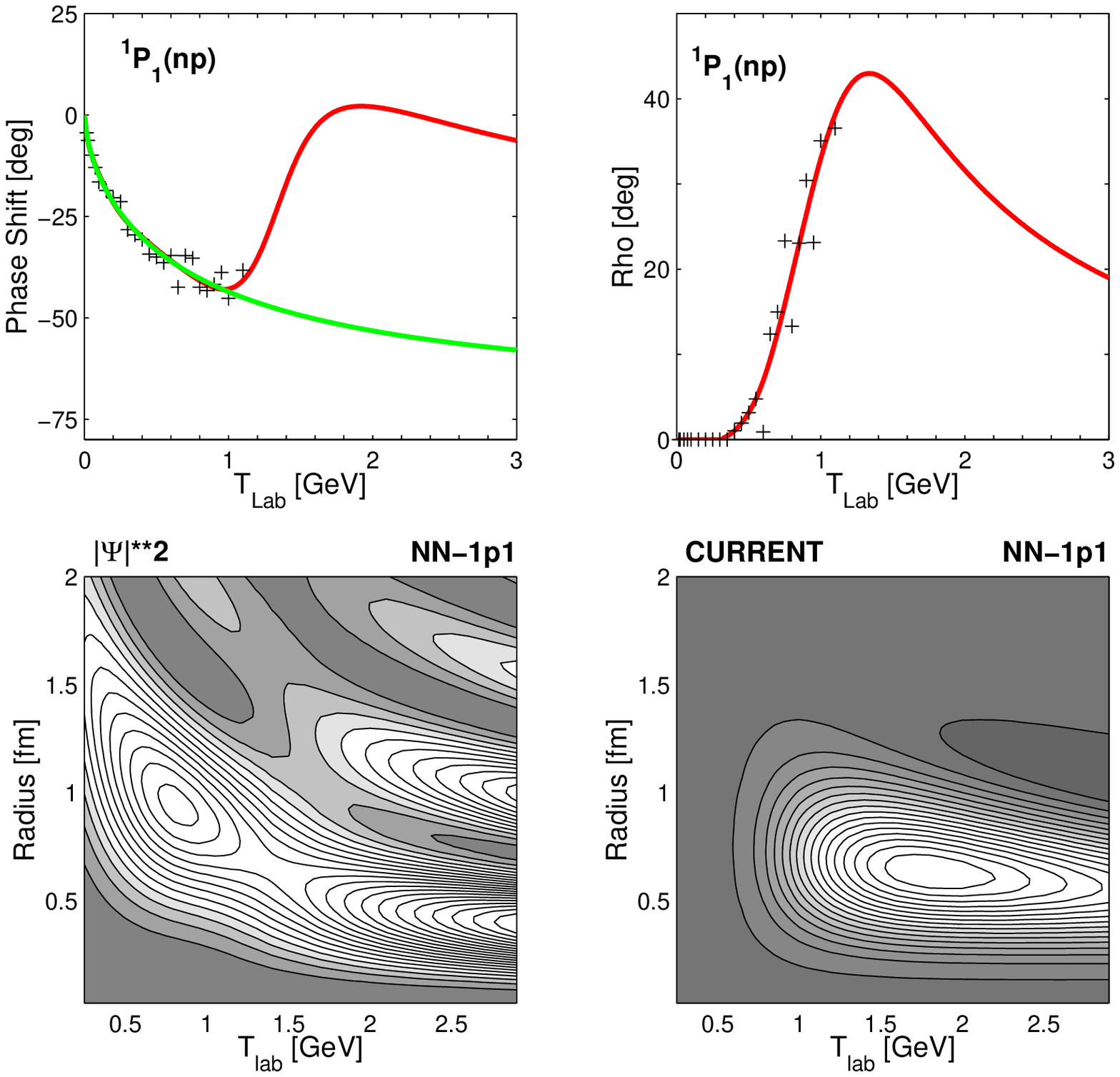,scale=.3}
\epsfig{file=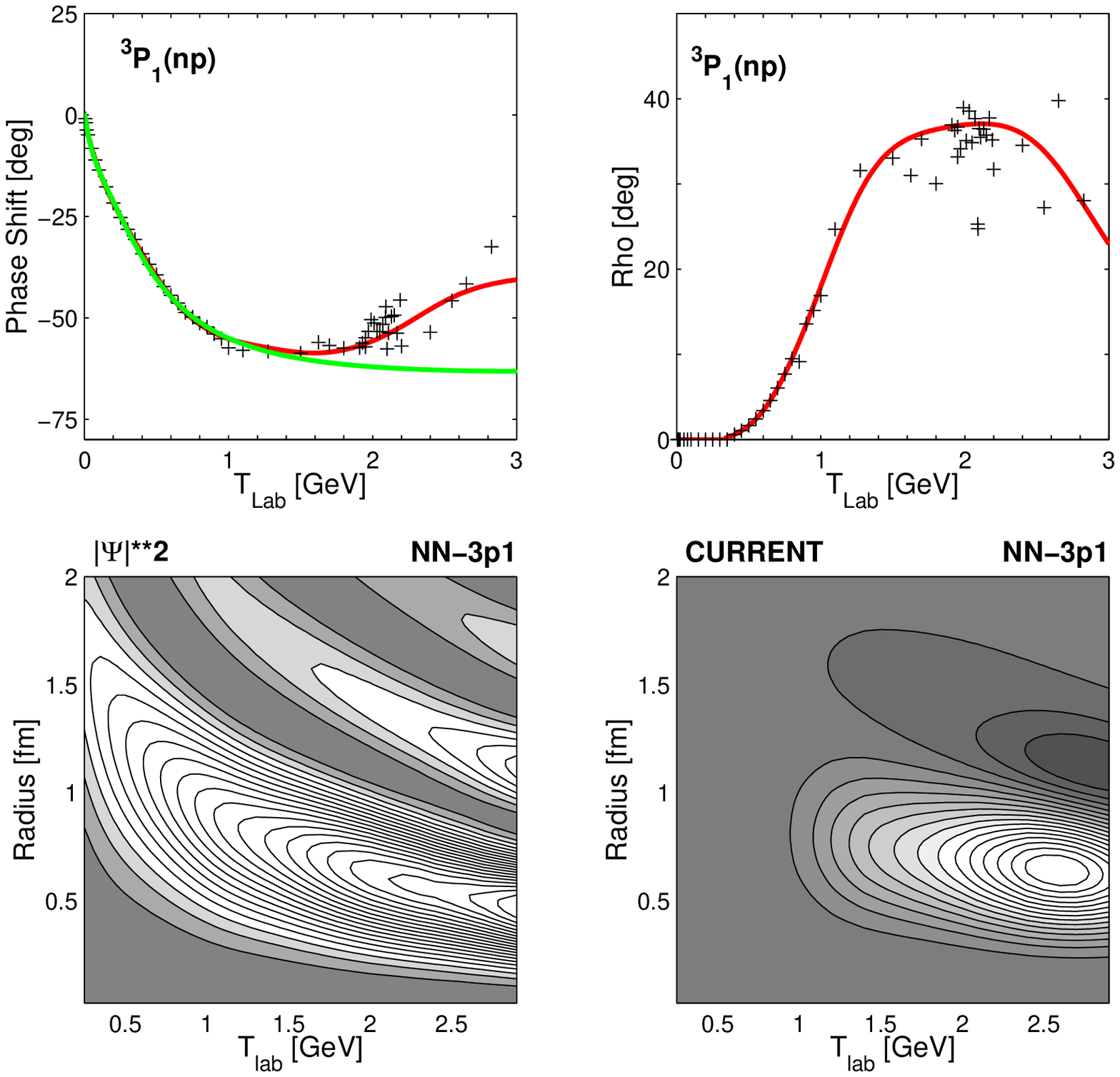,scale=.3}
\\
\epsfig{file=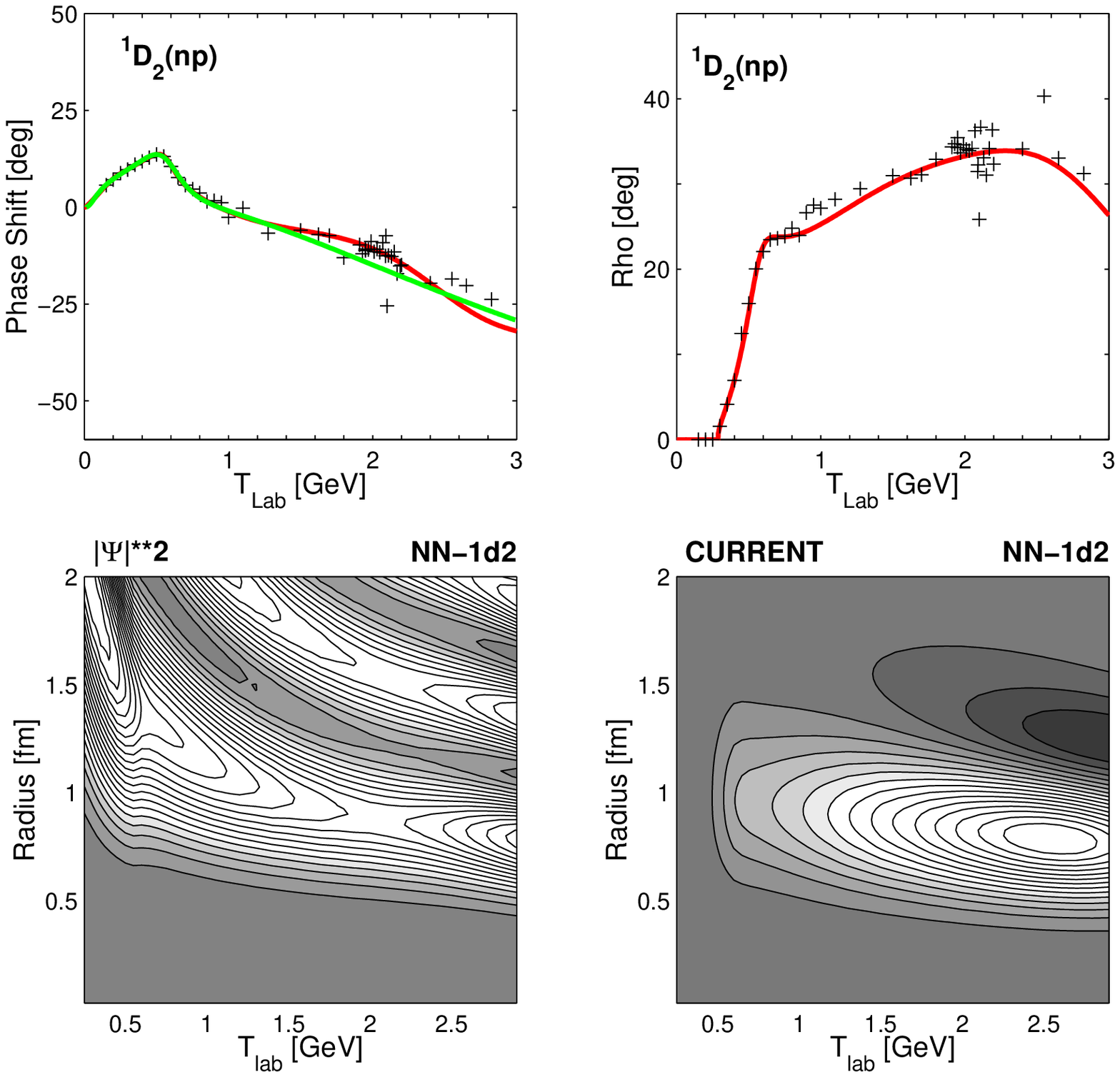,scale=.3}
\epsfig{file=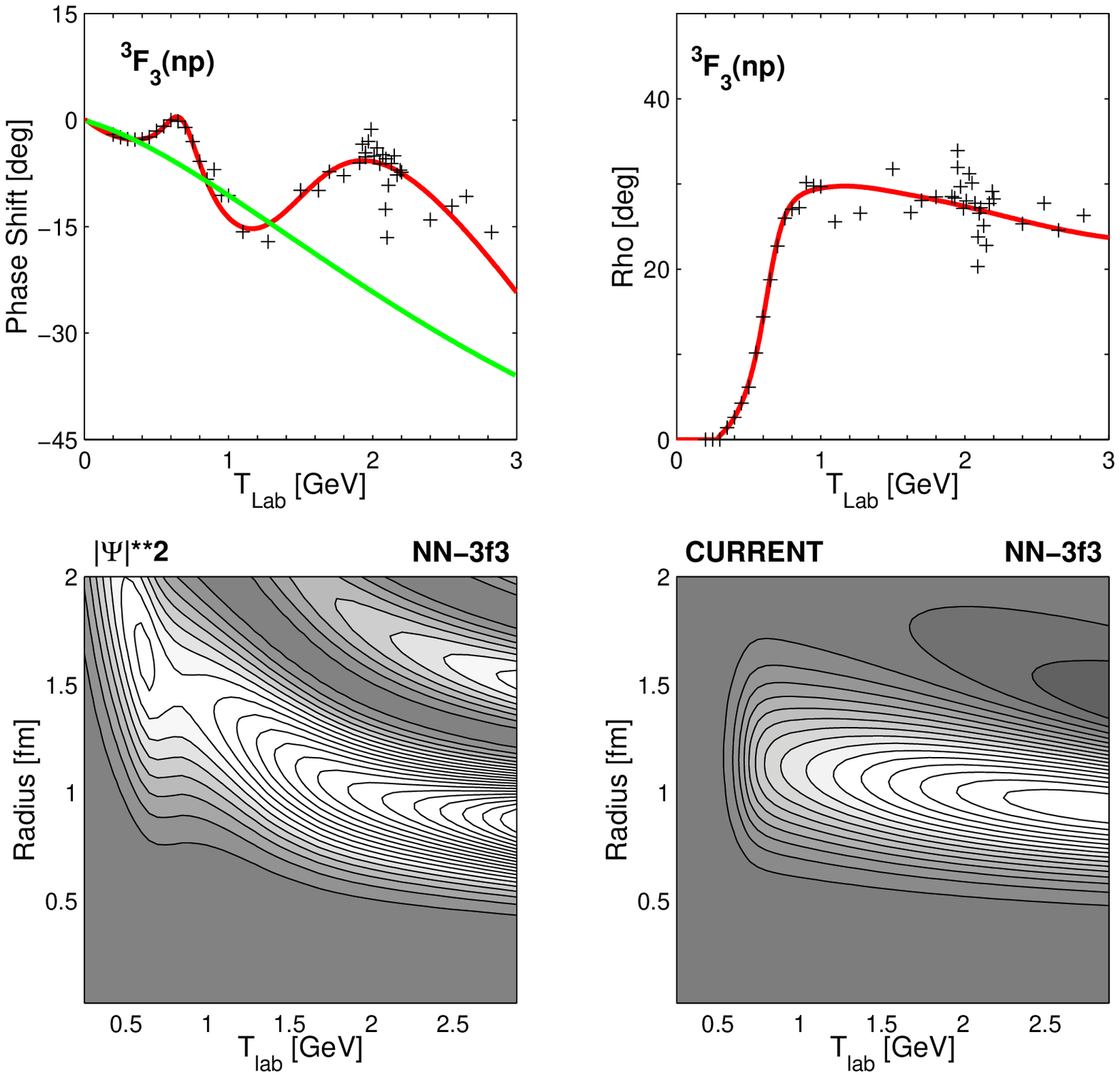,scale=.3}
\\               
\epsfig{file=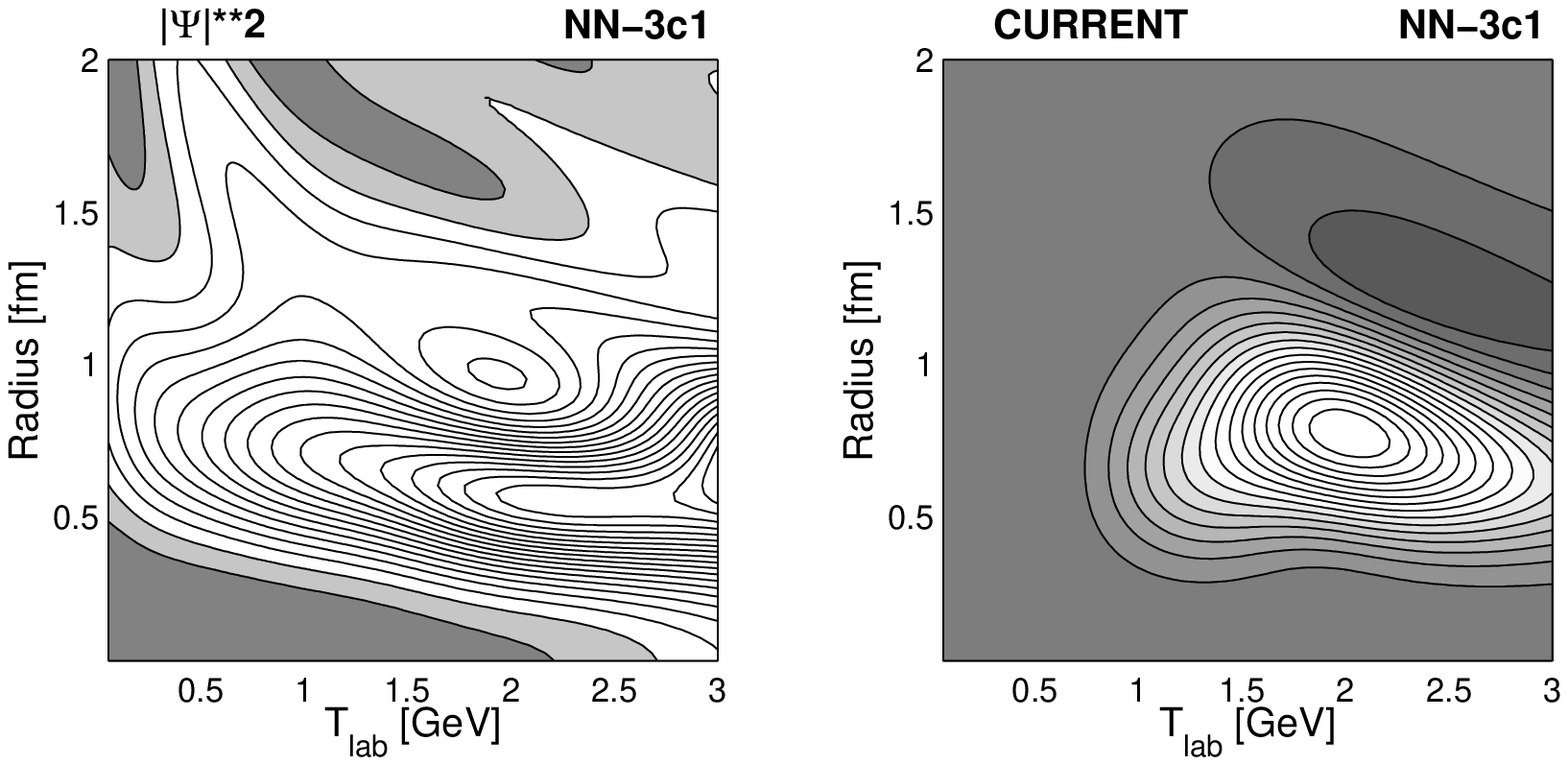,scale=.3}
\epsfig{file=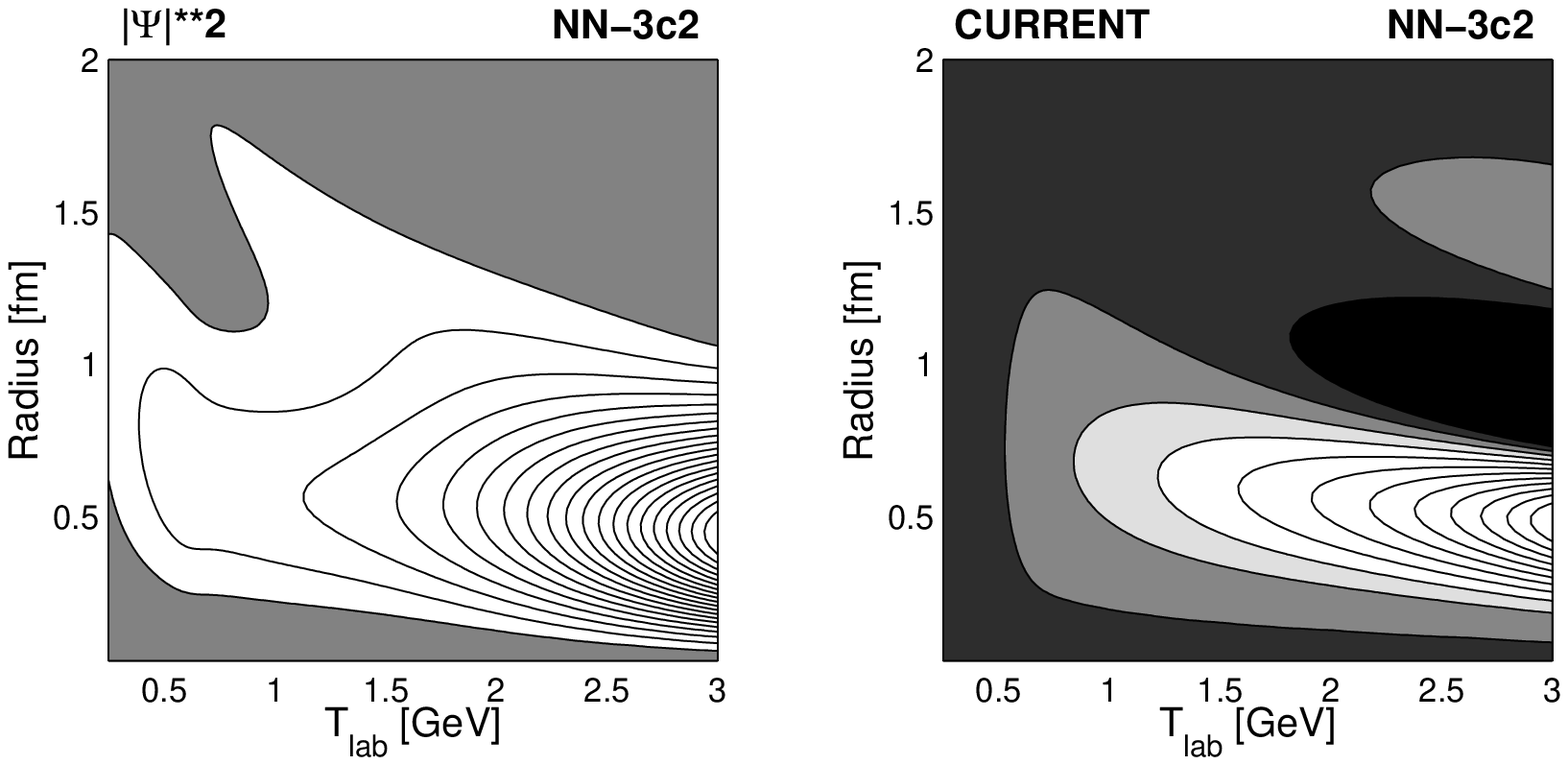,scale=.3}
\caption{Block matrices containing the reproduced $\delta$ and $\rho$ of FA00
\break  and energy/radial probabilities with loss of flux}
\label{ompresults}
\end{figure}

In Fig. \ref{ompresults} we show as a function of radius and energy the probability
density and the loss of flux. Four diagrams, arranged in a $2\times2$ matrix,
belong to one channel. Diagram-11 contains the real $\delta_\alpha$ 
(solid line for continuous solution
and crosses for single energy solutions), and the inversion potential, 
see Fig. \ref{ivspot}, reference
phases (shaded), diagram-12 contains the absorption angle $\rho_\alpha$ (the reference
 potential generates no absorption), diagram-21 shows probability surfaces as a function
of energy and radius, diagram-22 shows surfaces of loss of flux as a function
of energy and radius. For coupled channels we show in Fig. \ref{ompresults} the traces of 
probabilities and currents for the $^3SD_1$ and $^3PF_2$ channels.
 For the $T=0$ channels
only np data, $<1.2$ GeV, contribute and Arndt's solution $>1.2$ GeV is not determined
from data. This becomes relevant in $^1P_1$, $^3SD_1$. The figures support the
mental image claimed above in which nucleons do not fuse and their individuality
is maintained. The produced  mesons stem from one or the other nucleon and not
from a fused compound system. The strengths of separable potentials yield large tables,
are available on request, and shall be published elsewhere. They show the $\Delta$ resonance
and $N^*$ contributions in the relevant partial waves with the implication that all
dependencies are very smooth over many hundreds of MeV. None of the results support
the formation of intermediate fused NN systems, at least the phase shift solutions
do not support any such formation.

\section{Summary}

The primary purpose of this study was a comparison of the Paris, Nij-2, Reid93, AV18,
Nij-1, ESC96 and inversion potentials in the realm of an optical model extension using
boundary conditions and/or separable potentials. Applications are in progress for studies
of microscopic nucleon-nucleus scattering analysis in folding models, studies of exotic
nuclei and calculations needed in the technology of waste management with medium
energy particles. Despite these classical applications we seek collaboration with
groups pursuing QCD inspired theoretical development and calculations.

\end{document}